\documentclass[a4,12pt]{article}

\usepackage[utf8]{inputenc}

\usepackage{iftex}

\usepackage[sort&compress,numbers]{natbib} 
\usepackage[margin=2.5cm]{geometry}

\usepackage{amsmath}
\usepackage{amssymb}

\usepackage{graphicx} 
\graphicspath{ {./figures/} }
\usepackage[caption=false]{subfig}

\usepackage{dcolumn} 
\usepackage{bm} 

\usepackage{doi} 
\usepackage{hyperref} 

\DeclareMathOperator{\Tr}{Tr}
\DeclareMathOperator{\MSbar}{\mathrm{\overline{MS}}}

\newcommand\he[1]{#1^\dagger}
\newcommand\gr[1]{\mathrm{#1}}
\newcommand\SU[1]{\mathrm{\gr{SU(#1)}}} 
\ifPDFTeX
\newcommand\U[1]{\mathrm{\gr{U(#1)}}} 
\else
\providecommand{\U}{}
\renewcommand{\U}[1]{\mathrm{\gr{U(#1)}}} 
\fi

\newcommand{\nn}{\nonumber \\}

\begin{document}

\newcommand{\HEL}{\affiliation{%
		Department of Physics and Helsinki Institute of Physics,
		PL 64, FI-00014 University of Helsinki,
		Finland }}
		
\newcommand{\TSU}{\affiliation{%
		Tsung-Dao Lee Institute \& School of Physics and Astronomy, Shanghai Jiao Tong University, Shanghai 200240, China }}

\title{Infrared physics of the 3D $\mathrm{SU(2)}$ adjoint Higgs model at the crossover transition}

\author{Lauri Niemi,$^{a,b,}$\thanks{\href{mailto:lauri.b.niemi@helsinki.fi}
    {lauri.b.niemi@helsinki.fi}} \,
  Kari Rummukainen,$^{a,}$\thanks{\href{mailto:kari.rummukainen@helsinki.fi}
    {kari.rummukainen@helsinki.fi}} \,
  Riikka Seppä,$^{a,}$\thanks{\href{mailto:riikka.seppa@helsinki.fi}
    {riikka.seppa@helsinki.fi}} \,
  David J. Weir$^{a,}$\thanks{\href{mailto:david.weir@helsinki.fi}
    {david.weir@helsinki.fi}}
 
}

  \date{$^a$\small{\textit{Department of Physics and Helsinki Institute of Physics, \\
  		P.O. Box 64, FI-00014 University of Helsinki, Finland}} \\ 
  	$^b$\small{\textit{Tsung-Dao Lee Institute, Shanghai Jiao Tong University, Shanghai 200240, China}} \\[2ex]
  	\today
  }

  \begin{flushright}
    HIP-2022-18/TH
  \end{flushright}

  \begingroup
  \let\newpage\relax
  \maketitle
  \endgroup
  
  \begin{abstract}
    We study the crossover phase transition of the $\SU2$ Georgi-Glashow model in three dimensions. In this model, a confining condensate of topological 't\,Hooft-Polyakov monopoles exists in the Higgs regime. We use lattice Monte Carlo simulations to study the monopole gas across a crossover transition, and demonstrate that gradient flow can be used to renormalize the otherwise divergent monopole number density. 
	The condensation of the monopoles means that the theory admits also a massive photon-like excitation. We show that the renormalized monopole number density is approximately proportional to the square of the photon mass, in agreement with semiclassical results.
	Our results give insight into behaviour of the Higgs regime near crossover, which has boarder implications for beyond the Standard Model theories containing adjoint scalar fields.
  \end{abstract}

  \section{Introduction}
  
  An important class of gauge theories are those involving Higgs fields in the 
  adjoint representation, commonly known as Georgi-Glashow theories. They appear, among others, in grand unification schemes \cite{Georgi:1974sy} and as effective descriptions for hot Yang-Mills theories above the deconfinement temperature \cite{Appelquist:1981vg,Kajantie:1997tt}. The simplest case of $\SU2$ theory with a single adjoint Higgs field is conceptually interesting for the reason that the Higgs regime admits topologically stable defects, 't\,Hooft-Polyakov magnetic monopoles \cite{tHooft:1974kcl,Polyakov:1974ek}. Therefore, the $\SU2$ model provides a useful testing ground for studying strongly-coupled physics of magnetic monopoles in a simplified setting.

  Our focus will be on the $\SU2$ adjoint Higgs theory in three dimensions. The 3D case is well-known for allowing for a semiclassical description of confinement in which the Higgs field and 't\,Hooft-Polyakov monopoles play a crucial role.
  At the perturbative level, the Higgs mechanism effectively leaves an unbroken $\U1$ gauge symmetry with a massless photon. As argued first by Polyakov, this notion breaks down when  monopoles are taken into account, and the true infrared (IR) behavior resembles that of a dilute Coulomb gas of monopoles \cite{Polyakov:1976fu}. At long distances the screening of magnetic charges gives rise to confinement, and the photon-like excitation seen in perturbation theory is actually a pseudoscalar with mass (squared) proportional to the number density of monopoles. Although Polyakov's reasoning was based on semiclassical approximations, subsequent lattice studies support the existence of a confining monopole condensate \cite{Nadkarni:1989na,Hart:1996ac,Davis:2001mg}. The monopole interpretation loses its usefulness in the ``symmetric phase'' of the theory where the Higgs mechanism is lifted. Yet it is widely believed, and backed up by lattice simulations, that the symmetric and Higgs regimes are analytically connected. For sufficiently large values of the scalar self coupling, the transition between the two ``phases'' is of the crossover type, rather than a true phase transition \cite{Nadkarni:1989na,Hart:1996ac,Kajantie:1997tt,Davis:2001mg}.\footnote{This is not the case in $(3+1)$ dimensions, where the Higgs regime admits a massless photon and is thus separate from the gapped confining phase \cite{Lee:1985yi,Afferrante:2020hqe}.} 
  
  Semiclassical approximations are not justified in the vicinity of the crossover transition where strong coupling effects of the ``symmetric'' regime are important. The purpose of the present paper is to revisit the monopole gas picture with fresh lattice simulations. In the presence of fluctuations, the monopole number density $n$ is ultraviolet (UV) divergent because of short-lived monopole-antimonopole pairs. We demonstrate that a renormalized definition of $n$ with a well-defined continuum limit can be given by means of the gradient flow \cite{Luscher:2010iy}. The flow smooths the field configurations, leaving only well-separated monopoles that are responsible for photon mass generation in the semiclassical treatment. Our renormalized definition of monopole number density allows us to link semiclassical descriptions of the long-range physics of monopoles to a fully nonperturbative lattice simulation of the gauge-Higgs system. In particular, by measuring the photon mass $M_\gamma$ directly on the lattice, we demonstrate that the relation $M_\gamma^2 \propto n$ is reasonably accurate at the nonperturbative level and applicable also in the vicinity of the crossover.

  Remainder of this paper is organized as follows. We introduce the 3D model and summarize the known semiclassical results in section~\ref{sec:model-continuum} before discussing the lattice discretization in section~\ref{sec:lattice}. Section~\ref{sec:grad-flow} describes our implementation of the gradient flow on the gauge-Higgs system. Our results concerning the crossover transition, renormalized monopole number and the photon mass are presented in section~\ref{sec:results}. The concluding section~\ref{sec:conclusions} summarizes our main findings and discusses their implications. Some early results for this work originally appeared in the proceedings paper \cite{Niemi:2021ghk}.

  \section{The three-dimensional $\SU2$ Georgi-Glashow model}
  \label{sec:model-continuum}
  
  The $\SU2$ adjoint Higgs Lagrangian in three spatial dimensions is
  \begin{align}
  \label{eq:action-continuum}
  \mathcal{L} = \frac12 \Tr F_{ij} F_{ij} + \Tr D_i \phi D_i \phi + m^2_3 \Tr \phi^2 + \lambda_3 (\Tr \phi^2)^2 ,
  \end{align}  
  where $\phi = \frac12 \phi^a \sigma^a$, $\phi^a \in R$ and $\sigma^a$ are the Pauli matrices, and $D_i \phi = \partial_i
  \phi + ig_3 [A_\mu, \phi]$. The couplings $g_3$, $m_3^2$ and $\lambda_3$ have positive mass dimensions, hence $\mathcal{L}$ defines a super-renormalizable theory. 
  
  Although our goal is to better understand the role played by magnetic monopoles in the 3D model, let us briefly explain how the Lagrangian eq.\,(\ref{eq:action-continuum}) may arise as an effective description for physics in $(3+1)$ dimensions. At high temperature, the IR thermodynamics of pure $\SU2$ Yang-Mills is described by an adjoint Higgs theory of the above form \cite{Appelquist:1981vg}. The Higgs field then corresponds to the temporal gauge field $A_0$, which obtains mass of order $gT$ from Debye screening. This description is valid in the deconfined phase of $\SU2$ gauge theory \cite{Kajantie:1997tt}.
  Analogous dimensional reduction can be applied to a $(3+1)$D Georgi-Glashow model, integrating out the Debye screened field but keeping the low-energy modes of the original Higgs field.
  The resulting effective theory will, again, be of form eq.\,(\ref{eq:action-continuum}) and is valid for momenta $\ll gT$. This could be relevant for describing a cosmological phase transition in the early universe in models with electroweak triplet scalars \cite{Patel:2012pi,Niemi:2020hto}. Finally, a whole different application for 3D adjoint Higgs theories comes from holographic considerations of inflationary universes \cite{McFadden:2009fg}.

  Focusing now on the 3D case, the theory defined by eq.\,(\ref{eq:action-continuum}) has phase structure depending on two dimensionless parameters,
  \begin{equation}
  x = \frac{\lambda_3}{g_3^2}, \qquad y = \frac{m_3^2}{g_3^4}.
  \end{equation}
  At mean-field level, $y > 0$ corresponds to the confining phase and $y < 0$ to the Higgs phase where monopoles appear. The nonperturbative phase structure is known from lattice simulations \cite{Hart:1996ac, Kajantie:1997tt}. For small $x \lesssim 0.3$, the phases are separated by a first-order phase transition at a slightly positive $y$. At larger $x$, the phase transition disappears, and a smooth crossover takes place instead.\footnote{To be precise, refs.~\cite{Hart:1996ac,Kajantie:1997tt} considered local quantities only. Hence, it is not strictly ruled out that some nonlocal, e.g. topological, effect could turn the apparent crossover into a proper phase transition, like in the 3D Abelian Higgs model for instance \cite{Kajantie:1998zn}. In the present case, a crossover transition is nevertheless expected based on semiclassical reasoning. The need to consider nonlocal observables was emphasized in ref.~\cite{Davis:2001mg}, whose conclusions also support the crossover behavior.}

  One may apply semiclassical methods to study the IR behavior of the Higgs regime, leading to Polyakov's description in terms of a weakly-interacting monopole gas. In unitary gauge $\phi^a = v \delta_{a3}$, the Higgs mechanism produces two massive gauge fields with mass $m_W = g_3v$, while monopole condensation gives mass to the leftover ``photon''. Polyakov's results \cite{Polyakov:1976fu} for the monopole number density $n$ and the photon mass $M_\gamma$ are, up to constants of proportionality,
  \begin{align}
  \label{eq:n-semiclassical}
  n &\sim \frac{m_W^{7/2}}{g_3} \exp \Big[-\frac{4\pi m_W}{g_3^2} f(\lambda_3/g_3^2) \Big] \\
  \label{eq:photon-mass-semiclassical}
  M_\gamma^2 &\sim \frac{n}{\pi g_3^2},
  \end{align}
  where $f(x) \approx 1 + x/2$ for small $x$ \cite{Forgacs:2005vx}. The mass associated with a monopole is 
  \begin{align}
  \frac{M}{g_3^2} = \frac{4\pi m_W}{g^2_3} f(\lambda_3/g_3^2).
  \end{align}
  The semiclassical description is valid when the monopole gas is dilute, requiring $M/g_3^2 \gg 1$, or $v/g_3 = \sqrt{-y/x} \gg 1$. This is satisfied deep in the Higgs regime but not near $y \approx 0$ where the crossover takes place.

  \section{Lattice formulation}
  \label{sec:lattice}
  
  Our lattice action is
  \begin{align}
  \label{eq:lattice-action}
  S =& \; \beta \sum_{x, i<j} \Big(1 - \frac12 P_{ij}(x) \Big) + 2 a \sum_{x,i} \Big( \Tr \phi(x)^2 - \Tr \phi(x) U_i(x)\phi(x+i) \he U_i(x) \Big) \nn
  & + a^3 \sum_x \Big( m^2_L \Tr \phi^2 + \lambda_3 (\Tr \phi^2)^2 \Big) ,
  \end{align}
  where $a$ is the lattice spacing, $\beta = 4/(ag_3^2)$, $U_i(x)$ is an $\gr{SU(2)}$ gauge link and $P_{ij}(x) = U_i(x) U_j(x+i) \he U_i(x+j) \he U_j(x)$ is the Wilson plaquette. Because of super-renormalizability, only the mass parameter requires renormalization whereas $\lambda_3$ and $g_3^2$ can be taken equal to continuum values. Consequently, complete lattice-continuum relations can be found analytically \cite{Laine:1995np,Laine:1997dy}. For $m_3^2(\bar{\mu})$ given in continuum $\MSbar$ scheme, one has, up to corrections that vanish linearly in small $a$,
  \begin{align}
  \label{eq:lattice-mass}
  m_L^2 &= m_3^2 - \frac{\Sigma }{4 \pi a}\Big( 4g_3^2 + 5\lambda_3 \Big) - \frac{1}{16\pi^2} \Big[ \Big( 20\lambda_3 g_3^2 - 10 \lambda_3^2 \Big)\Big( \ln\frac{6}{a \bar{\mu}} + \zeta \Big) + 20\lambda_3 g_3^2 \Big( \frac14 \Sigma^2 - \delta \Big) \nn
  & \quad +2 g_3^4 \Big( \frac54 \Sigma^2 + \frac{\pi}{3}\Sigma - 8\delta - 8\rho + 4\kappa_1 - 2\kappa_4 \Big) \Big].
  \end{align}
  Here $\bar{\mu}$ is the $\MSbar$ scale, which we take equal to $g_3^2$, and the numerical constants $\Sigma, \zeta, \rho, \delta, \kappa_1, \kappa_4$ are given in \cite{Laine:1997dy}.\footnote{Partially $O(a)$ improved lattice-continuum relations are presented in \cite{Moore:1997np}.  We did not implement this, because the $O(a)$ contribution to eq.\,(\ref{eq:lattice-mass}) is lacking and this turns out to be the most significant part of the improvement.  However, see \cite{Moore:2019lua} for a non-perturbative determination of the full improvement in a related case.}

  In the following we take the continuum $x,y$ parameters as fixed inputs, and treat the lattice action with the mass parameter fixed according to eq.\,(\ref{eq:lattice-mass}) as an approximation to the continuum theory. Lattice cutoff effects are small provided that $\beta \geq 6$, as indicated by good scaling behavior of mass ratios in the lattice model~\cite{Hart:1996ac}.

  Discussion of magnetic monopoles in lattice regularization requires special care as topology is not well defined on the lattice. We follow ref.~\cite{Davis:2000kv} in constructing a  discrete magnetic field in three spatial dimensions. We define the projector
  \begin{align}
  \Pi_+(x) = \frac12 \left(\mathbf{1} + \frac{\phi}{\sqrt{2\Tr\phi^2}}\right)
  \end{align}
  and modified link variables through
  \begin{align}
  u_i(x) = \Pi_+(x) U_i(x) \Pi_+(x+i),
  \end{align}
  which correspond to links of an Abelian gauge symmetry.
  The Abelian field strength tensor $\alpha_{ij}$ and the associated lattice magnetic field $\hat{B}_i$ is obtained from elementary plaquettes constructed from $u_i(x)$:
  \begin{align}
  \hat{B}_i(x) &= \frac12 \epsilon_{ijk} \alpha_{jk},
  \end{align}
  \begin{align}
  \alpha_{ij} &= \frac{2}{\hat{g}_3} \arg \Tr u_i(x) u_j(x+i)\he u_i(x+j) \he u_j(x).
  \end{align}
  Here $\hat{g}_3 = 2/\sqrt{\beta}$ is the gauge coupling in natural lattice units. It is straightforward to check that $\hat{B}_i$ is a gauge-invariant quantity. 
  
  The magnetic charge density inside a unit cube is given by the divergence
  \begin{align}
  \label{eq:charge-density}
  \rho_M(x) = \sum_{i=1}^{3} [\hat{B}_i(x+i) - \hat{B}_i(x)],
  \end{align}
  and is quantized in units of $4\pi / \hat{g}_3$.\footnote{In our simulations, the probability of finding more than one unit of charge in a unit cube was much less than $1\%$.}
  The total charge density, obtained by summing eq.\,(\ref{eq:charge-density}) over all lattice sites, vanishes identically on a periodic lattice.\footnote{Frequently in the monopole literature, so-called twisted C-periodic boundary conditions are imposed, allowing non-vanishing magnetic charge to exist on the lattice while also preserving translational invariance \cite{Kronfeld:1990qu,Edwards:2009bw}. This is useful e.g. for measuring the mass associated with a monopole \cite{Davis:2000kv,Davis:2001mg,Rajantie:2005hi}. Our lattices in this work use standard periodic boundary conditions and so a magnetic monopole will always be accompanied by an antimonopole.} 
  In contrast, the combined number density of monopoles and antimonopoles, 
  \begin{align}
  \label{eq:number-density}
  n = \frac{1}{a^3 N_s^3} \frac{\hat{g}_3}{4\pi} \sum_x |\rho_M(x)|
  \end{align}
  is generally non-zero for a given field configuration. Here $N_s$ is the number of lattice sites in each direction and $V = (aN_s)^3$ is the physical volume. For $y$ above the crossover point, i.e. in the ``confining'' regime, the magnetic monopoles become indistinguishable from the vacuum and their interpretation as particles is lost. The measurement of $n$ as described here can nevertheless be carried out.
 
  The quantity defined by eq.\,(\ref{eq:number-density}) is badly UV divergent: field fluctuations can create monopole-antimonopole pairs at short distances, and the number of such pairs increases without limit as the lattice spacing is reduced. At long distances they look like dipoles and can produce only short-ranged magnetic fields. Hence, these ``UV pairs'' should be inessential for correlations over long distances, and one expects only widely separated monopoles to affect physics in the IR. This is assumed also in the semiclassical treatment \cite{Polyakov:1976fu}, but in the presence of fluctuations, counting only monopoles with separation $\gg a$ requires systematic renormalization of the number density eq.\,(\ref{eq:number-density}). We turn to this in the next section.

  \section{Gradient flow for the gauge-Higgs system} 
  \label{sec:grad-flow}
  
  Gradient flow refers to a flow in field space towards stationary points of the action along the direction of steepest descent. It is a convenient tool for producing smooth, renormalized fields from configurations generated by a Markov chain simulation \cite{Luscher:2010iy}.
  To perform gradient flow, we promote the fields to be functions of a new dimensionless ``time'' $\tau$. Writing the links as $U_i(x) = \exp \left[i \theta_i^a(x) \sigma^a  / 2 \right]$, our gradient flow is defined by the equations
  \begin{align}
  \label{eq:grad-flow-1}
  \frac{\partial U_i(x)}{\partial\tau} &= -\frac12 i a g_3^2 \sigma^a \frac{\partial S}{\partial \theta_i^a(x)} U_i(x) \\
  \label{eq:grad-flow-2}
  a\frac{\partial \phi^a(x)}{\partial \tau} &= - \frac{\partial S}{\partial \phi^a(x)},
  \end{align}
  where the dependence on $\tau$ in the fields is implicit. Expressions for the gradients are given in Appendix~\ref{sec:grad-flow-appendix}. 
  
  \begin{figure}[t]
  	\centering
  	\subfloat[$\xi = 0.69a$]{
  		\includegraphics[trim=257 96 154 346,
  		clip,width=.32\textwidth]{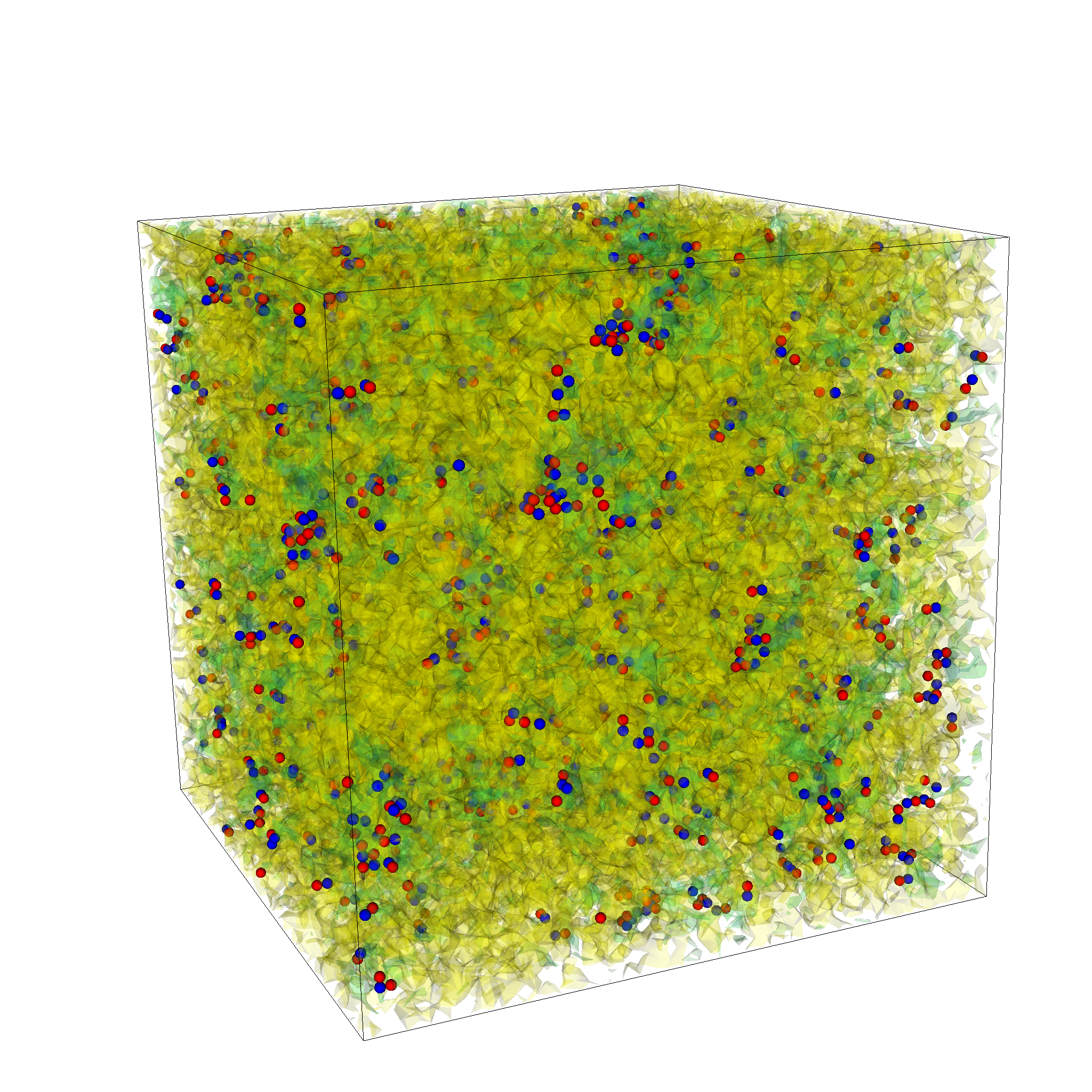}
  	}\hfill
  	\subfloat[$\xi = 1.55a$]{\includegraphics[trim=257 96 154 346,
  		clip,width=.32\textwidth]{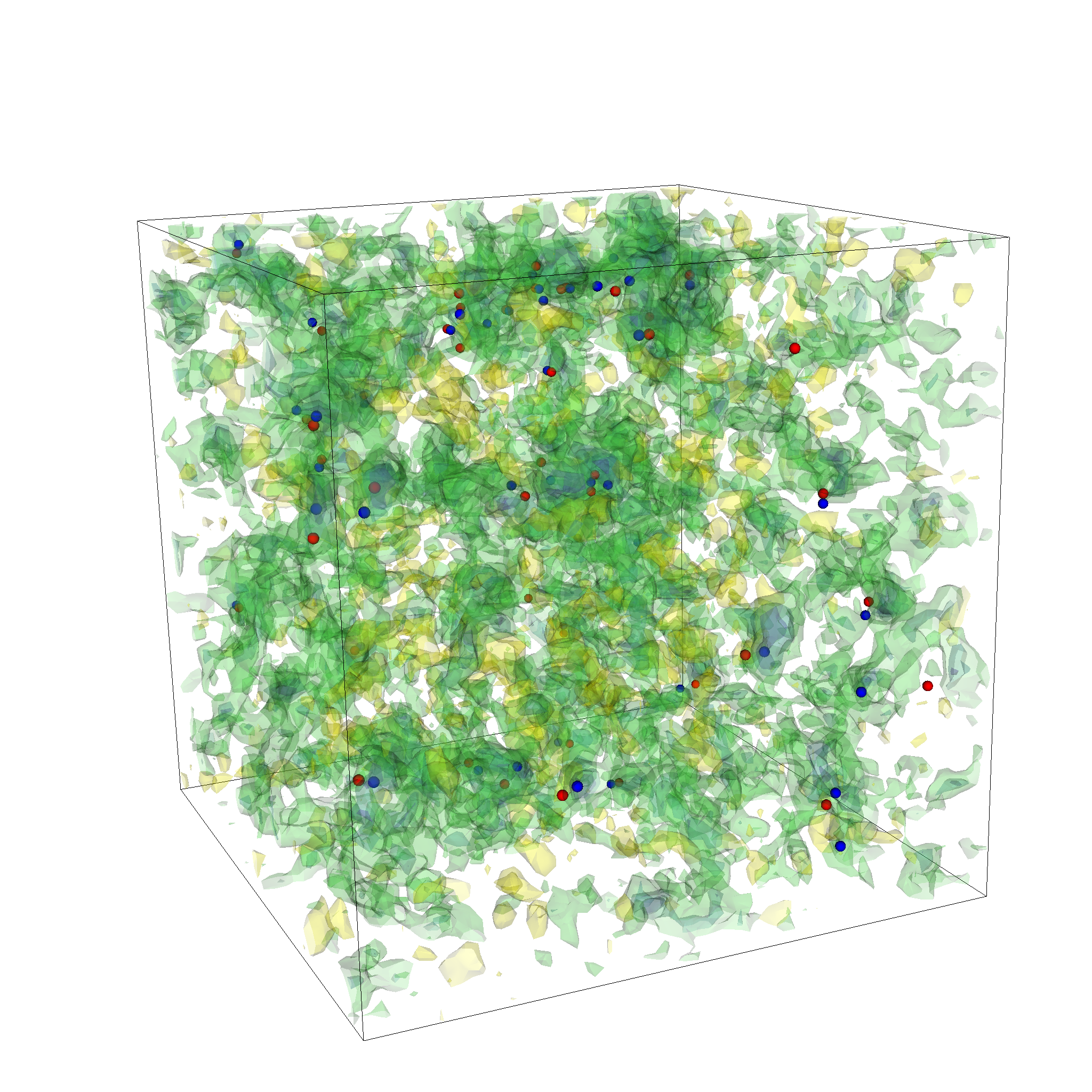}
  	}\hfill
  	\subfloat[$\xi = 3.46a$]{\includegraphics[trim=257 96 154 346,
  		clip,width=.32\textwidth]{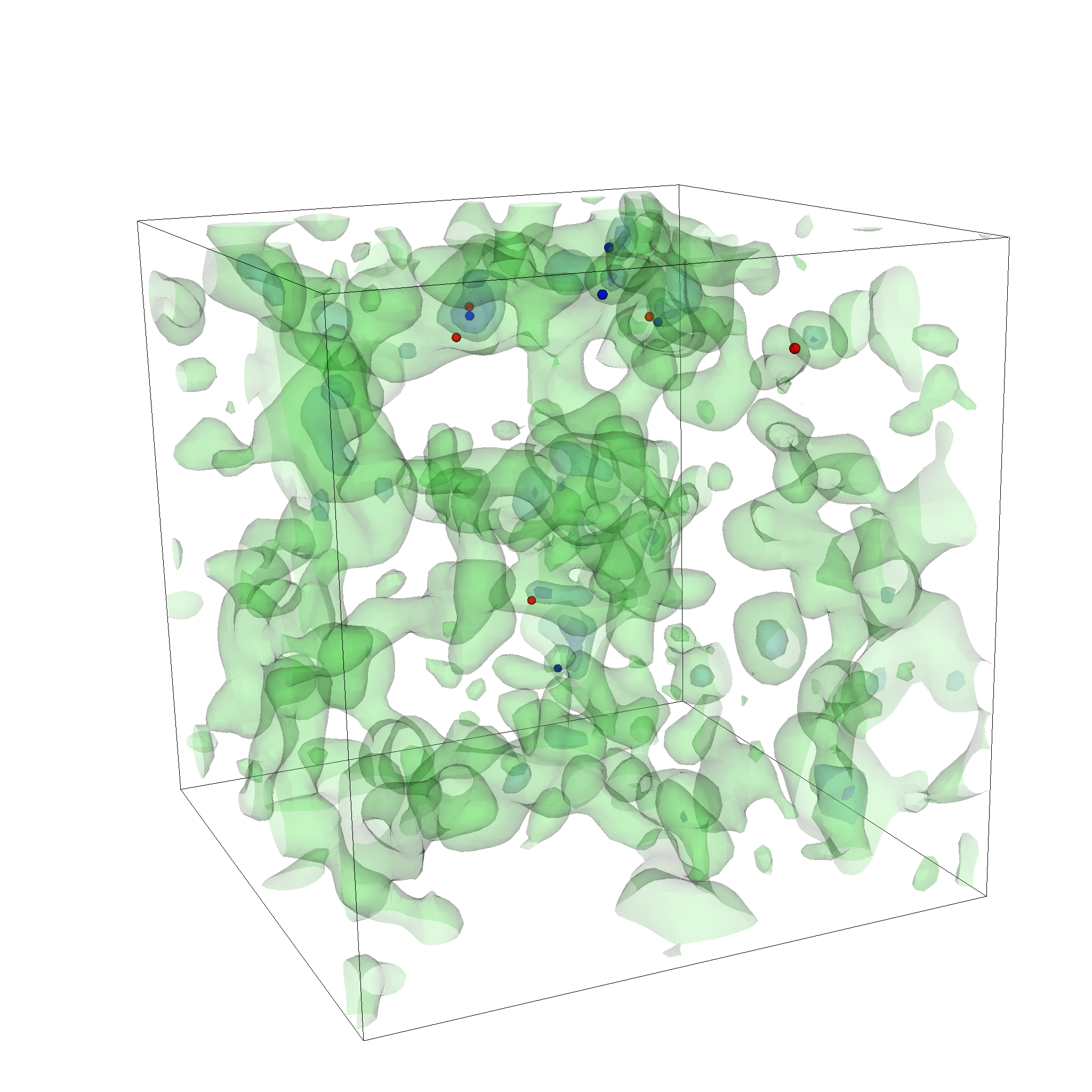}
  	}
  	\caption{Gradient flow applied to a hot configuration containing 
  		't\,Hooft-Polyakov monopoles and antimonopoles (blue and red dots, respectively).
  		Isosurfaces of $\Tr\phi^2$ are shown in shades of green, and $\xi$ labels the 
  		smoothing radius as described in the text.
  		Monopoles appear where the Higgs field approaches zero (bluer regions).
  		Monopole-antimonopole pairs produced by short-distance fluctuations
  		annihilate as the flow proceeds, and only widely separated monopoles remain at long flow times.
  		The simulation is for $x=0.35$, $y=0.03$, $\beta = 8$ on a $64^3$ lattice. 
  		Full movie is available at  \url{https://zenodo.org/record/6375749}.
  		}
  	\label{fig:snapshots}
  \end{figure}

  In three dimensions, the flow smooths gauge field configurations in a spherical region of radius $\xi = \sqrt{6\tau}a$. The smoothing facilitates the annihilation of monopole-antimonopole pairs separated by distances $\lesssim \xi$. This is illustrated in figure~\ref{fig:snapshots}, which shows snapshots of the system during gradient flow. Thus the monopole number density $n(\xi)$, measured from a smoothed configuration, is a renormalized quantity and should have a finite continuum limit. Here $\xi$ labels the renormalization point.

  The 't\,Hooft-Polyakov monopoles are nontrivial configurations of both the gauge and Higgs fields, and it is crucial that the flow is applied simultaneously to both fields. Here a technical complication arises due to the presence of mass divergences. 
  The lattice mass parameter $m_L^2$ requires a negative counterterm, as shown in eq.\,(\ref{eq:lattice-mass}), to preserve the connection to continuum physics. However, the gradient flow smooths out UV fluctuations, reducing the need for the counterterm. Thus the Higgs potential, and in particular its minimum, changes during the flow, unless the mass is gradually modified to compensate for the change in UV fluctuations. It is not clear how this could be implemented in practice.
  
  The above problem has been mentioned previously in the context of electroweak sphaleron rate in \cite{Moore:1998swa}. There it could be circumvented by defining the observable in terms of the gauge fields only, whereas for 't\,Hooft-Polyakov monopoles the Higgs field is irreplaceable. Thus we must choose how to define the parameter $m_L^2 = m_L^2(\xi)$ when gradient flow is applied. In a sense this is simply a choice of renormalization scheme, and a good scheme for our purposes is one that gives well-behaved results for the monopole number density. 
  
  \begin{figure}[t]
  	\centering
  	\subfloat[]{
  		\includegraphics[width=.48\textwidth]{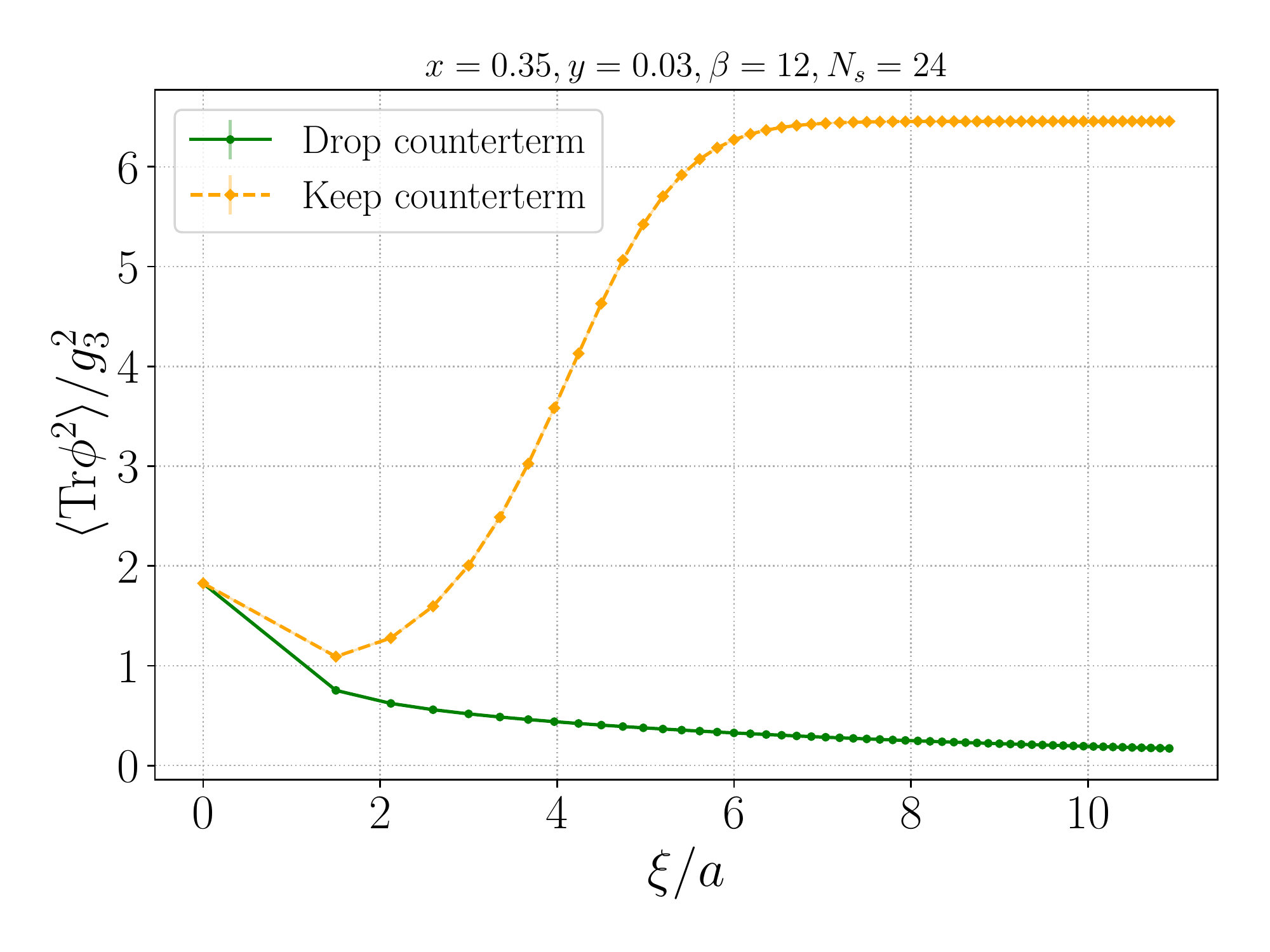}
  		\label{fig:test-ct-a}
  	}\hfill
  	\subfloat[]{
  		\includegraphics[width=.48\textwidth]{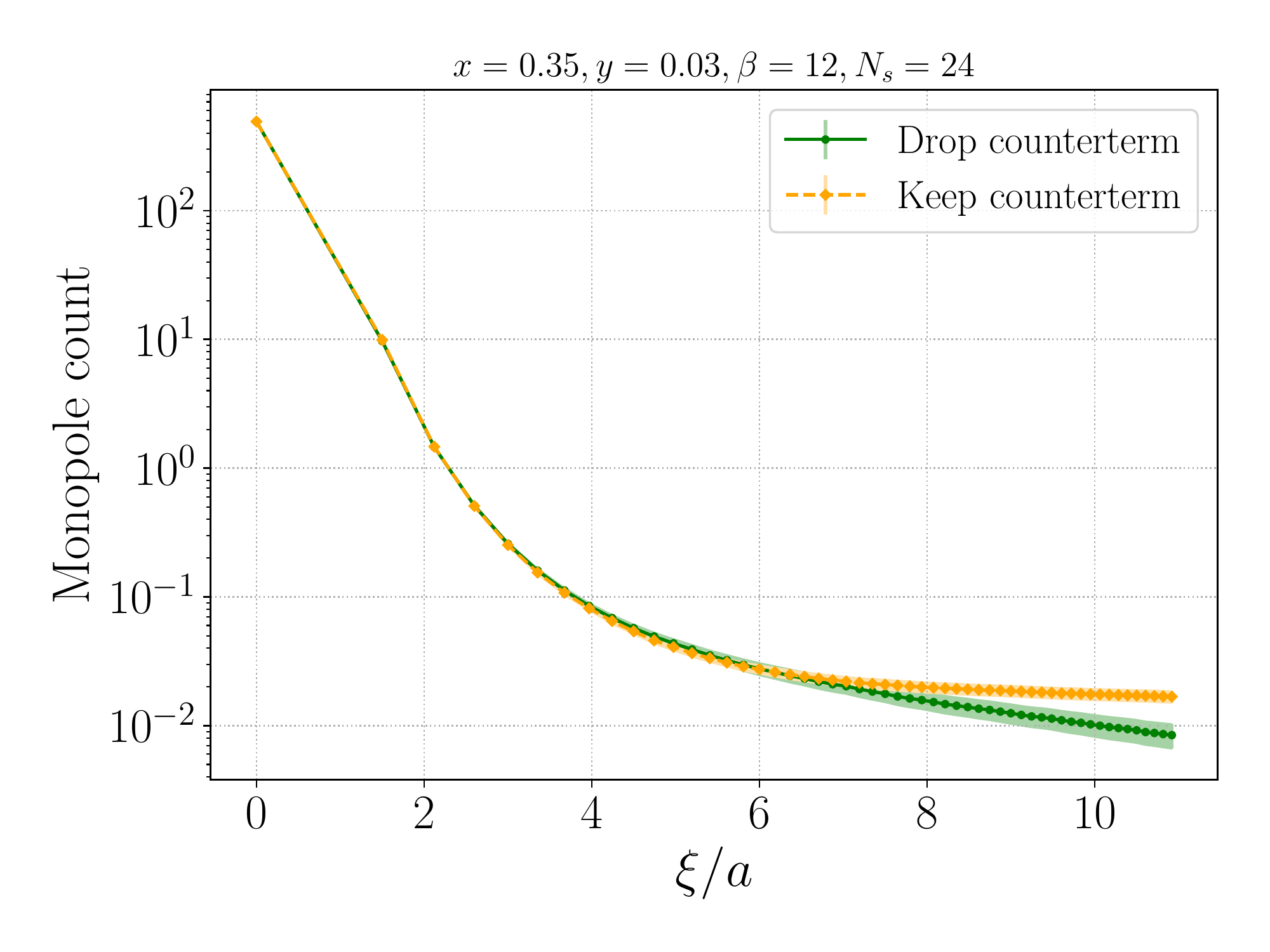}
  		\label{fig:test-ct-b}
  	}
  	\caption{Comparison of two the gradient flow schemes described in the text. (a) Higgs condensate, (b) average number of magnetic monopoles in the system. The bare mass $m_L^2$ appearing in the flow equations is larger (smaller) for solid (dashed) lines due to the presence or absence of a UV counterterm. Data points are shown for clarity; the spacing is nonuniform due to the nonlinear relation between $\xi$ and the flow time $\tau$.
  	}
  	\label{fig:test-ct}
  \end{figure}
  
  A simple choice would be to simply keep the counterterm. The flow will then drive the Higgs field towards its equilibrium value at smaller (more negative) $y$, as the subtracted counterterm is ``too large'' once UV fluctuations have been smoothed. This is illustrated by the dashed line in fig.~\ref{fig:test-ct-a}. However, for long gradient flows the counterterm should become irrelevant, and it is perhaps more intuitive to drop it altogether, \textit{i.e.} use $m_L^2(\xi) = y g_3^4$ in the gradient flow equations. This choice leads to a relatively smaller deviation from initial Higgs field values, but the system is driven towards the symmetric phase by the flow (solid line in the figure). For the number of monopoles these two schemes differ only at long flow times, as shown in the right-hand figure. Monopole-antimonopole annihilation is dictated by the cooling radius $\xi$ and is not strongly affected, at short time scales, by the Higgs field, which relaxes towards its classical vacuum on a slower time scale.

  In the following sections (and for figure \ref{fig:snapshots} above) we choose the latter option: the mass counterterm is dropped as we initiate the gradient flow. It should be emphasized that this choice of scheme affects only quantities renormalized with the gradient flow. The ensemble of initial ($\xi=0$) configurations is still generated using the action eq.\,(\ref{eq:lattice-action}) with $m_L^2$ given by eq.\,(\ref{eq:lattice-mass}).

  \section{Simulation results}
  \label{sec:results}

  Our simulation uses heatbath and over-relaxation algorithms to generate field configurations according to the canonical ensemble. The standard heatbath algorithm for $\SU2$ \cite{Kennedy:1985nu} requires a minor modification due to the Higgs hopping term, which is quadratic in the gauge link. We perform a separate accept/reject step to account for this interaction. For the Higgs, our update sweep consists of five over-relaxation updates using the algorithm described in ref.\,\cite{Kajantie:1995kf}, combined with a Metropolis update for ergodicity. 
  
  For the numerical analysis we concentrate on $x=0.35$, for which the phase transition is of the crossover type according to ref.~\cite{Kajantie:1997tt}. We have reproduced this result below.
  
  \subsection{The crossover transition}
  
\begin{figure}[t]
  	\centering
  	\subfloat[]{
  		\includegraphics[width=.480\textwidth]{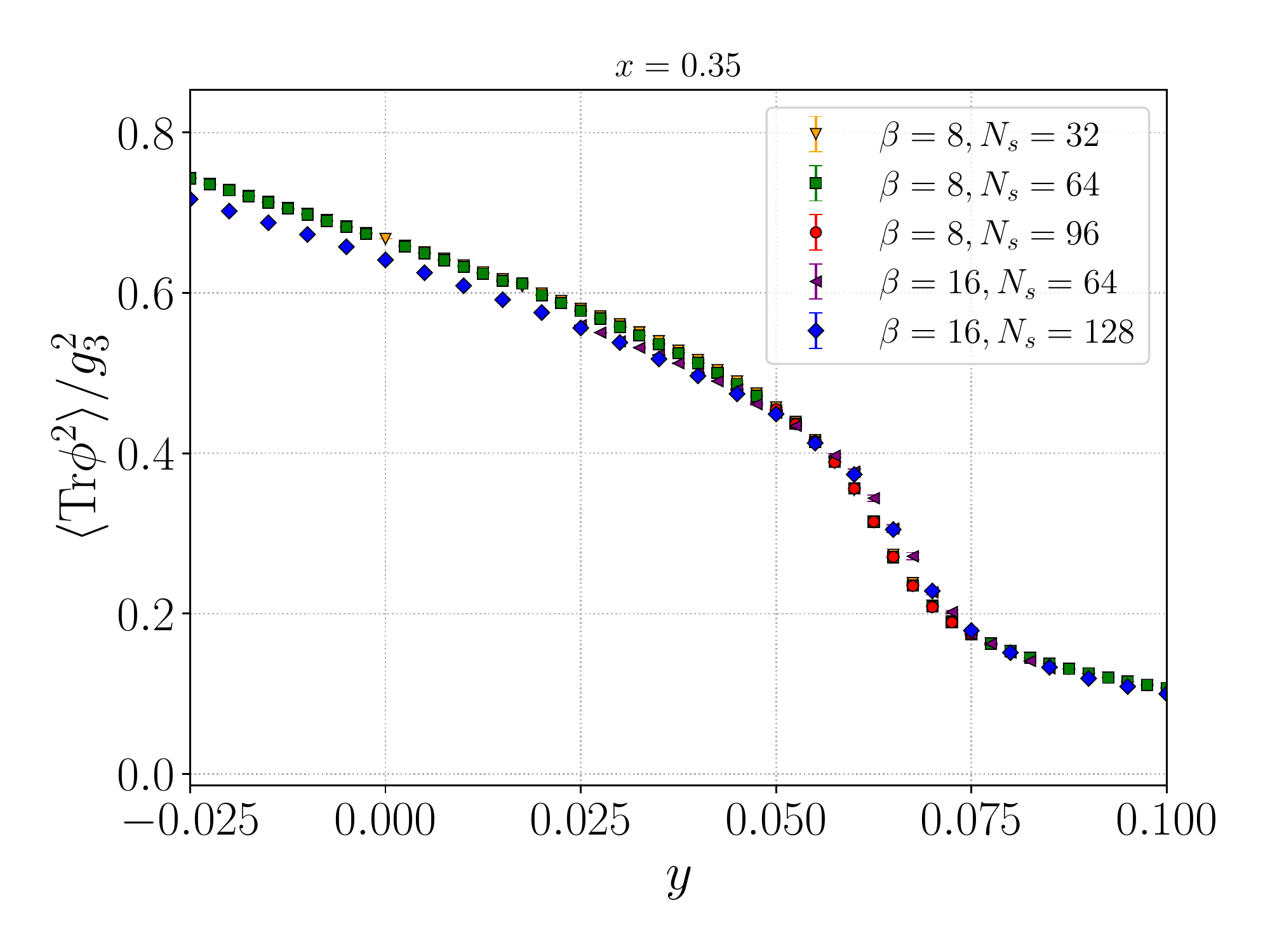}
  	}\hfill
  	\subfloat[] {
  		\includegraphics[width=.480\textwidth]{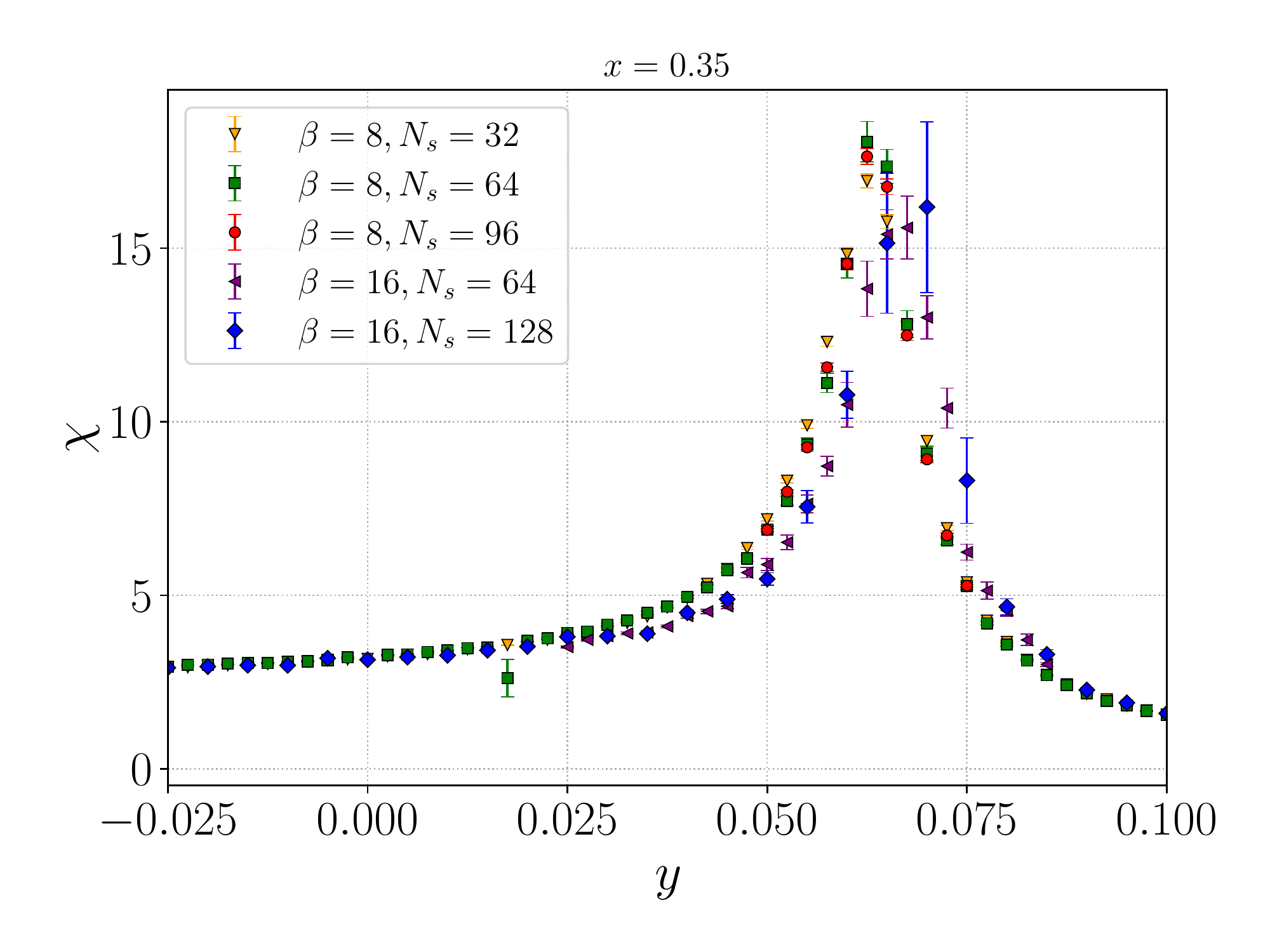}
  	}
  	\caption{(a) Quadratic Higgs condensate at different lattice spacings $a g_3^2 = 4/\beta$, converted to $\MSbar$ scheme. (b) Volume and $\beta$ dependence of the susceptibility (\ref{eq:susc}).}
  	\label{fig:Higgs-condensate}
  \end{figure}

  The phase transition point can be identified by using the Higgs condensate $\langle \Tr \phi^2\rangle$ as an effective order parameter. Gradient flow is not needed for this measurement. We show the dependence of $\langle \Tr \phi^2\rangle$ on $y$ in the left-hand panel of fig.~\ref{fig:Higgs-condensate}. Although the bare condensate is UV divergent, the divergence can be removed by an additive counterterm \cite{Laine:1995np}, giving the condensate in continuum $\MSbar$ scheme (scale $\bar\mu = g_3^2$).
  The sharp increase of the condensate between $0.05 < y < 0.075$ can be taken to indicate transition from a confinement-like vacuum state at large $y$ to Higgs-dominated behavior at smaller $y$. 
  
  The right-hand panel shows the dimensionless susceptibility 
  \begin{align}
  	\label{eq:susc}
  	\chi = V g_3^2 \left\langle \left( \Tr\phi^2 - \langle \Tr\phi^2 \rangle \right)^2 \right\rangle,
  \end{align}
  which peaks at the pseudocritical value $y_c \approx 0.065$. The precise value carries mild dependence on $\beta$ but is in agreement with ref.~\cite{Kajantie:1997tt}. The susceptibility is continuous and shows no power-like increase with the volume, consistent with crossover behavior.

  \subsection{Renormalized monopole number}
  \label{sec:results-n}

    \begin{figure*}[t]
    	\centering
    	\subfloat[]{
    		\includegraphics[width=.480\textwidth]{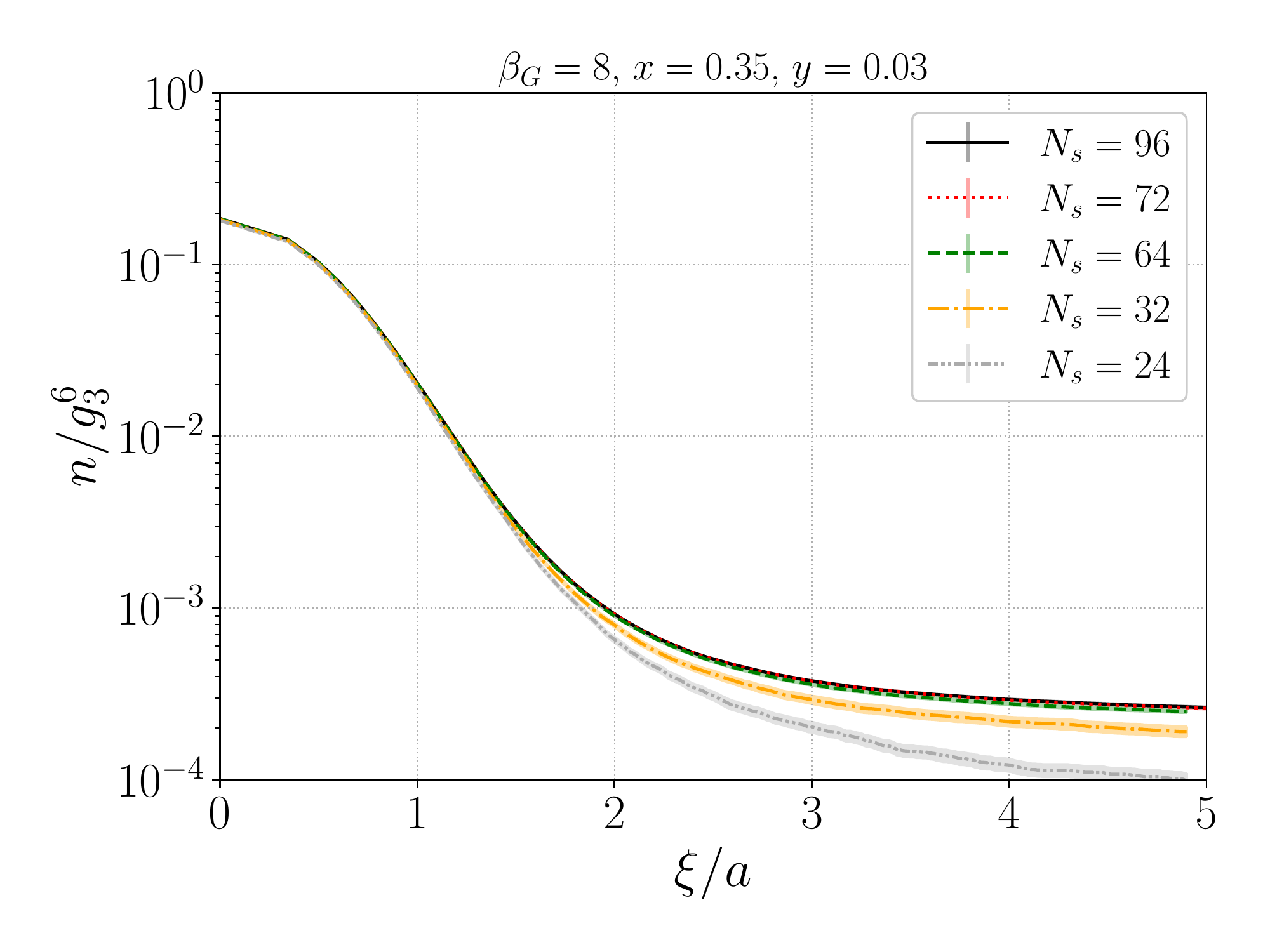}
    		\label{fig:number-density-a}
    	}\hfill
    	\subfloat[] {
    		\includegraphics[width=.480\textwidth]{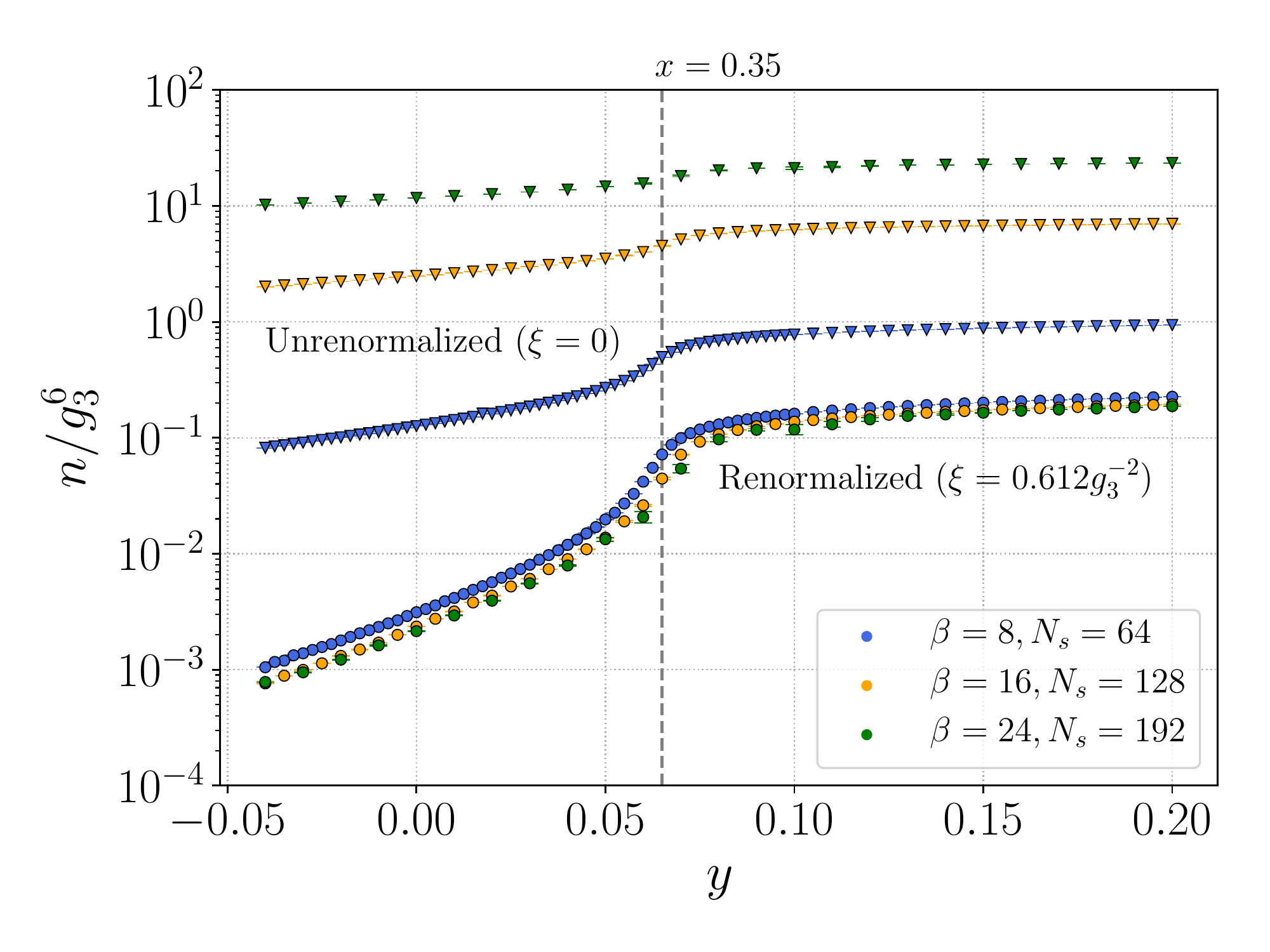}
    		\label{fig:number-density-b}
    	}
    	\caption{(a) Evolution of the monopole number density during gradient flow at different volumes and fixed $\beta=8$. (b) Renormalized and unrenormalized number densities as functions of $y$, at different values of $\beta$. The dashed line indicates where the crossover transition takes place.}
    	\label{fig:number-density}
    \end{figure*}

  For counting 't\,Hooft-Polyakov monopoles, we generate a canonically-distributed sample of field configurations and integrate the gradient flow equations separately for each configuration. The renormalized monopole number density $n(\xi)$ is then measured as described in section~\ref{sec:lattice}. We show its evolution with the smoothing radius $\xi$ in fig.~\ref{fig:number-density-a}. As is evident from the figure, the practical infinite volume limit for the monopole number at large $\xi$ is reached only on rather large lattices. This can be understood semiclassically \cite{Davis:2001mg}: For $y \ll y_c$, the monopole gas is dilute and their mean separation $D$ is large. To fully capture monopole screening and their condensation, the lattice side length should be $L \gg D$ in physical units. For instance, the large volume results for $n$ at $\beta=8$ seem to saturate towards $n/g_3^6 \approx 2.2\times 10^{-4}$. This gives a mean separation of $D g_3^2  \approx 16.5$. For $\beta=8$ this implies that the screening should be apparent on lattices with $N_s \gg 33$, in agreement with fig.~\ref{fig:number-density-a}. These estimates are $y$-dependent, as monopole configurations become rarer with decreasing $y$. 
  
  In fig.~\ref{fig:number-density-b} we show our results for the renormalized number density at different $y$; here the dashed line at $y = 0.065$ denotes the approximate crossover point. The lattices are large enough to allow at least partial (if not complete) screening of monopoles.\footnote{A similar plot was presented in \cite{Niemi:2021ghk} but with data from considerably smaller lattices. Although a clear continuum limit for the renormalized monopole number density was observed already in that case, the screening experienced by monopoles was likely only partial at best due to the small volumes used.} The renormalization point is fixed to $\xi = 0.612g_3^{-2}$. This provides enough smoothing that the most short-distance monopole pairs are annihilated, and the resulting number density shows no significant dependence on $\beta$, demonstrating that our renormalized definition of $n$ admits a well-behaved continuum limit. For comparison, the upper data points in fig.~\ref{fig:number-density-b} correspond to unrenormalized $\xi=0$ values for which no continuum limit is observed.
  
  Our choice of $\xi = 0.612g_3^{-2}$ is sufficient for obtaining a continuum limit for the number density, but in principle any larger $\xi$ should be a valid choice as well. We note, however, that more smoothing is not particularly useful as having too few monopoles remaining is bad for statistics: For example, in our $\beta=24$ simulation at $y=0.0$ and $\xi = 1.94g_3^{-2}$, most of our smoothed configurations did not contain even a single monopole pair. Thus the computational effort required for a reliable measurement of $n$ is substantially larger at long flow times.

  \subsection{Photon screening mass}

  For photon mass measurements we use the gauge-invariant vector operator $h_i = \epsilon_{ijk} \phi^a F_{jk}^a$ (in continuum notation). This is similar to the pure photon operator constructed by 't\,Hooft \cite{tHooft:1974kcl} but less noisy \cite{Kajantie:1997tt}. Our lattice implementation of $h_i$ calculates the field-strength $F_{ij}(x)$ from a clover-shaped Wilson loop; the full expression is given by eq.~(11) in ref.~\cite{Bonnet:2001rc}. This has reduced discretization errors compared to the ``naive'' expression involving only a single plaquette.

    \begin{figure*}[t]
  	\centering
  	\subfloat[]{
  		\includegraphics[width=.480\textwidth]{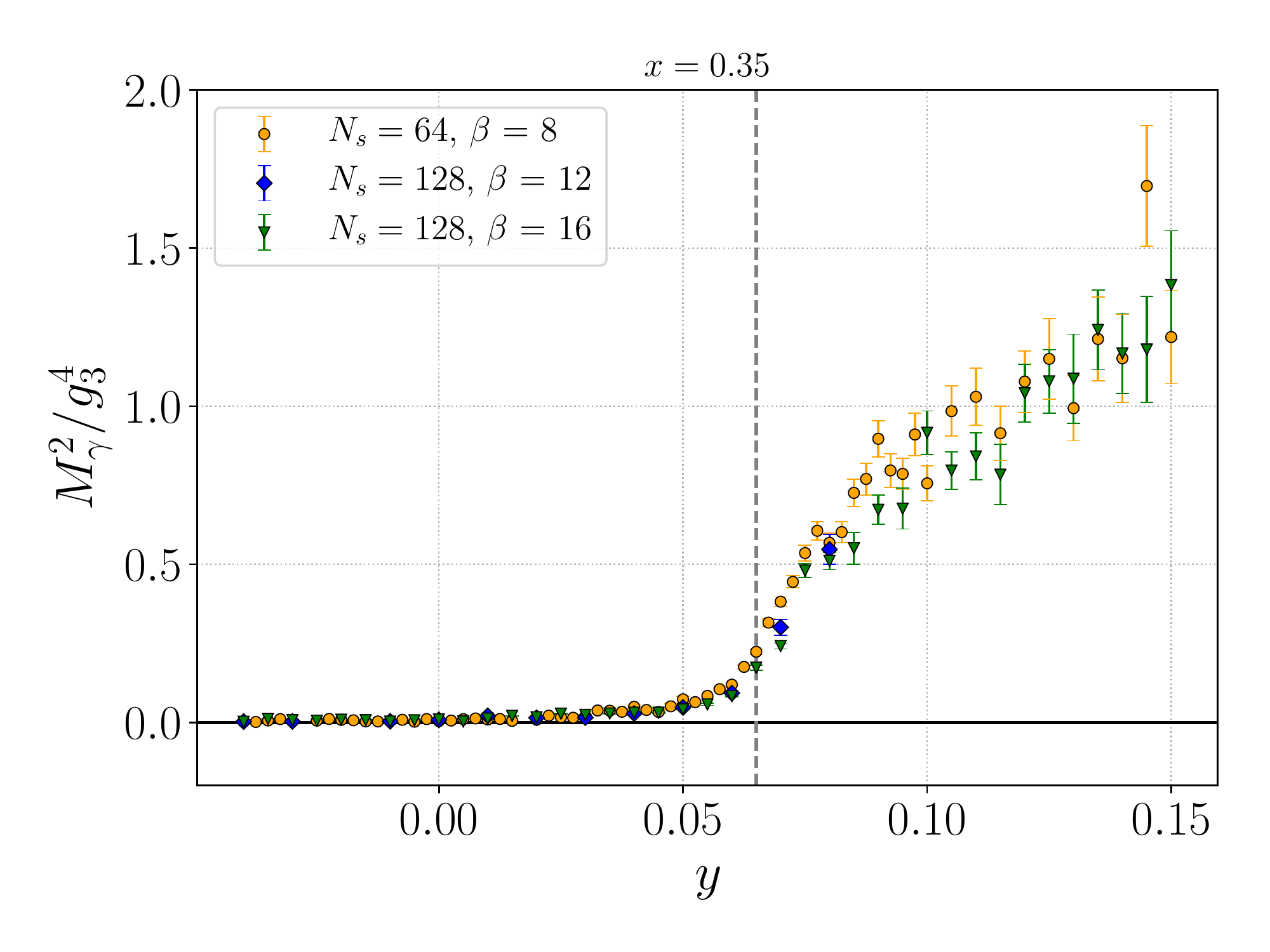}
  		\label{fig:msq-a}
  	}\hfill
  	\subfloat[] {
  		\includegraphics[width=.480\textwidth]{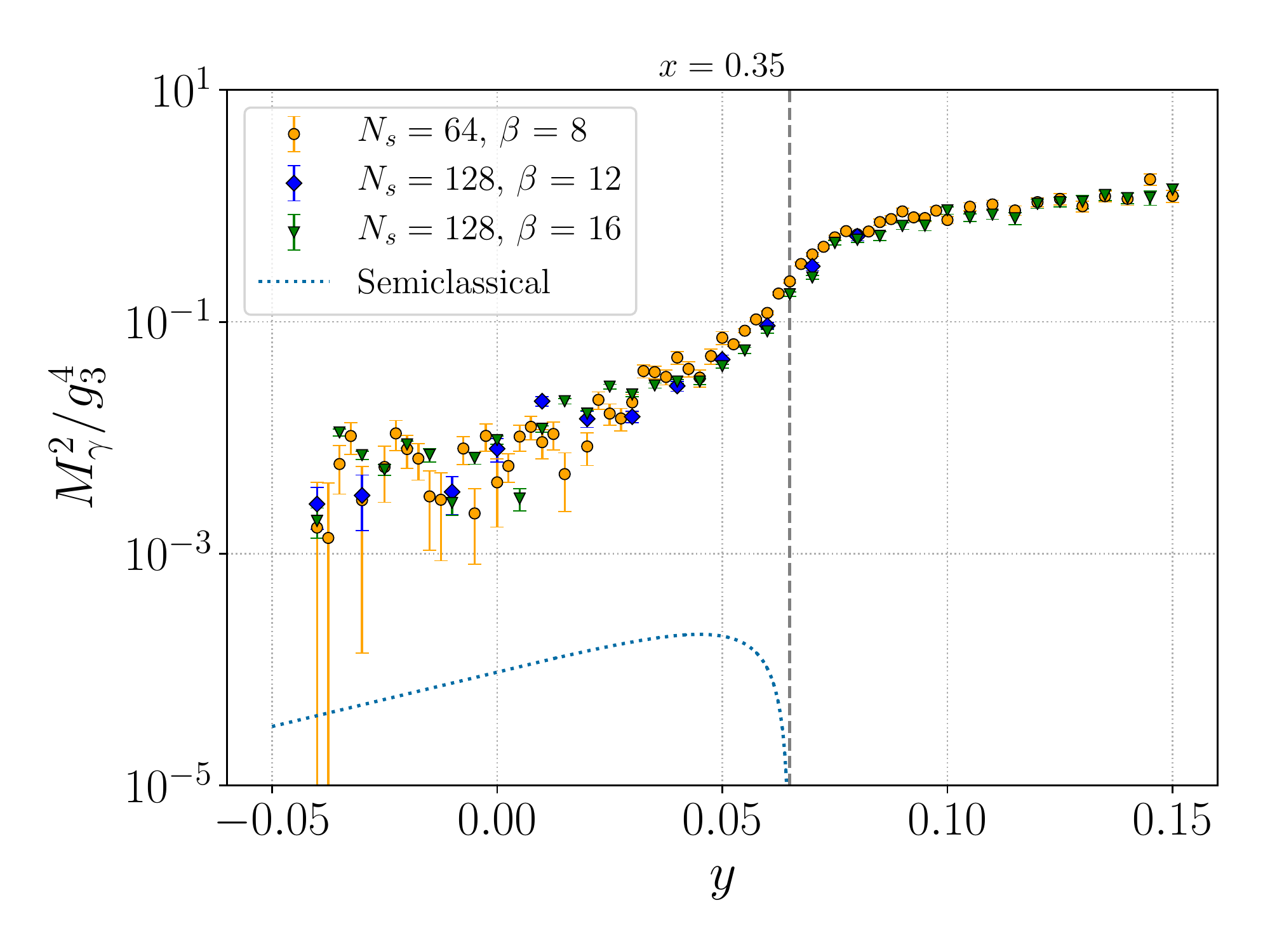}
  		\label{fig:msq-b}
  	}
  	\caption{(a): Square of the photon mass in physical units. The vertical dashed line at $y=0.065$ shows the approximate crossover point. (b): Same as (a) but on a logarithmic scale and showing also the semiclassical expectation, eq.\,(\ref{eq:photon-mass-semiclassical}), with undetermined proportionality constant.
  	}
  	\label{fig:msq}
  \end{figure*}
  
  \begin{figure}[t]
  	\centering
  	\includegraphics[width=.70\textwidth]{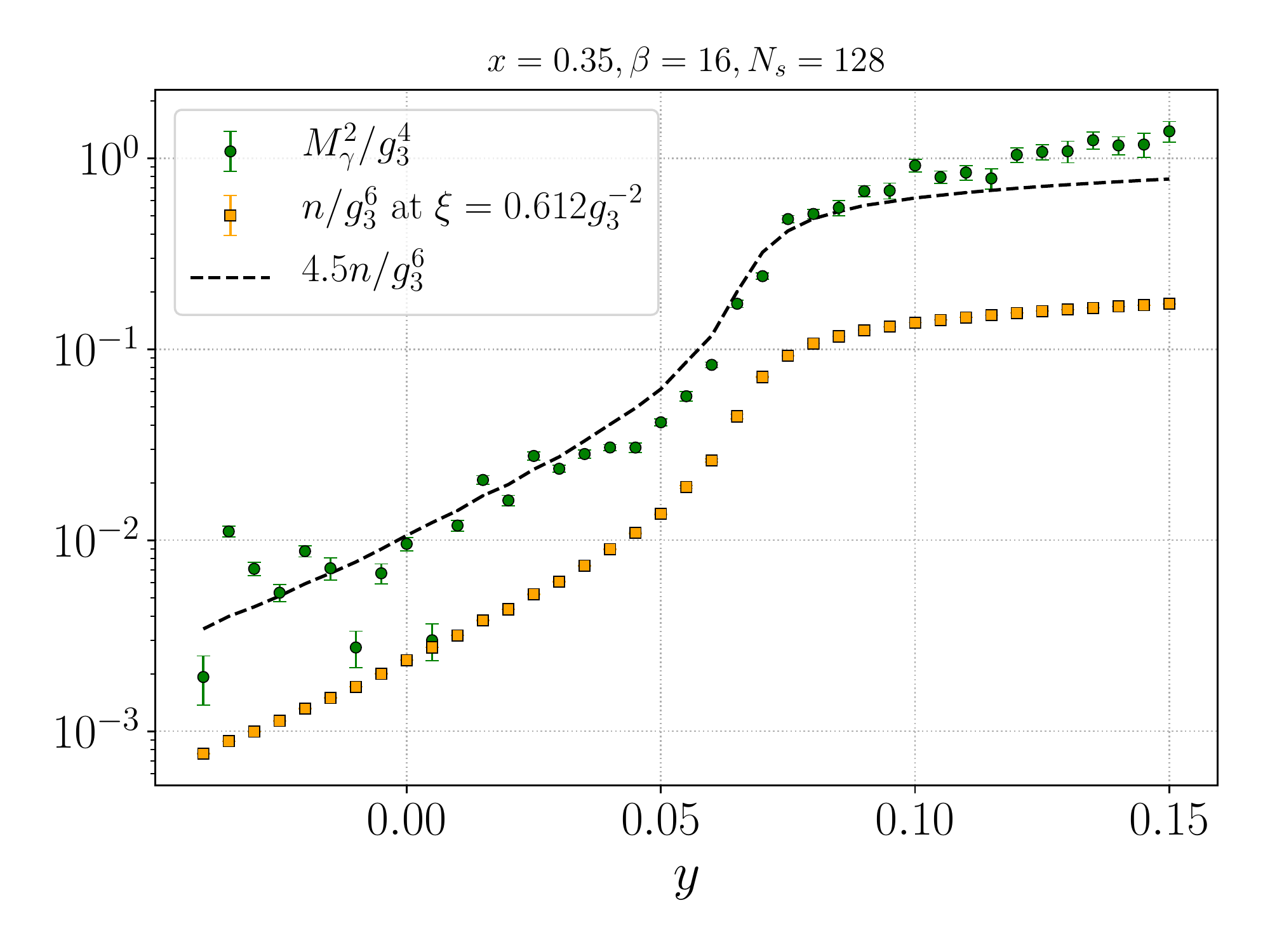}
  	\caption{Photon mass squared and the renormalized monopole number density, in appropriate units of $g_3^2$, on the same log-scale. The dashed line illustrates approximate proportionality between the two.}
  	\label{fig:msq-n-relation}
  \end{figure}

  The mass $M_\gamma$ is extracted from the plane-plane correlator of $h_3(x)$ along the $x_3$ direction. In practice we employ non-zero momentum in the transverse planes, i.e. the lattice operator is 
  \begin{align}
  \label{eq:photon-operator}
  O_3(z) = \frac{1}{N_s^2} \sum_{x_1, x_2} \delta_{x_3, z} \phi^a(x) F_{12}^a(x) e^{i2\pi x_1/N_s}
  \end{align}
  for the first momentum channel. This is useful because the photon mass is very small deep in the Higgs regime and so the zero-momentum correlator may need to be quite long before the mass can be extracted. The correlator asymptotes to 
  \begin{align}
  \langle O_3(z) O_3(0) \rangle \propto \left( e^{-\hat{E} z} + e^{-\hat{E}(N_s - z)}\right)
  \end{align}
  with $\hat{E} = \sqrt{(2\pi/N_s)^2 + (aM_\gamma)^2}$. We use the continuum dispersion relation instead of the full lattice expression; the difference is dwarfed by statistical uncertainty in the results because the momentum is small.
  
  Details of the measurement are as described in ref.~\cite{Kajantie:1997tt}. In particular, we improve the signal by performing a blocking transformation on the fields according to eqs.~(5.19)-(5.21) in \cite{Kajantie:1997tt}. This transformation replaces gauge links in the transverse directions by longer, appropriately smeared links, effectively increasing the transverse lattice spacing. Similar blocking is applied to the Higgs field. This reduces UV noise in the correlator measurement while preserving the signal. Results below are obtained after three subsequent blocking transformations and using the first non-zero momentum channel. We note that already the first blocking level has considerably smaller error bars compared to the unblocked case and that results from different blocking levels agree within statistical errors. The second momentum channel gives similar results as the first, but the correlator is noisier at long distances.

  We have measured $M_\gamma^2$ across the crossover transition; the results are shown in fig.~\ref{fig:msq}. We have data for $\beta=8, 16$ at fixed physical volume and at a larger volume for $\beta = 12$. The mass is comparable to $g_3^2$ above and near the crossover point $y_c \approx 0.065$, denoted by a dashed line in the figures. For smaller $y$, the mass is exponentially small as shown in the log-scale plot \ref{fig:msq-b}; the semiclassical result eq.\,(\ref{eq:photon-mass-semiclassical}) is plotted for illustration (shifted to start at $y=y_c$ instead of $y=0$). One sees that a proportionality constant of order $10^{2}$ is needed in eq.\,(\ref{eq:photon-mass-semiclassical}) to match the nonperturbative mass. Semiclassical intuition is not applicable at $y \geq y_c$, where the operator in eq.\,(\ref{eq:photon-operator}) presumably describes a confining gauge-Higgs state and obtains its mass from strong dynamics. Nevertheless, the mass $M_\gamma$ interpolates smoothly between the Higgs regime at small $y$ and the strongly coupled large-$y$ region, just as one would expect in the absence of a phase transition.

  At the lower end of our $y$ range the data is not accurate enough to convincingly give a nonzero value for the mass. This is visible as increasingly large error bars in the log-plot. The errors have been obtained with the jackknife method. In few cases the correlator fit gives a negative value for $M_\gamma^2$ and these points have been dropped from the figures.

  To address the question of whether the semiclassical proportionality $M_\gamma^2 \propto n/g_3^2$ holds nonperturbatively, we show in fig.~\ref{fig:msq-n-relation} the renormalized number density together with our $\beta=16$ results for the photon mass. The dashed line demonstrates approximate proportionality between the two quantities; however, the constant of proportionality naturally depends on the amount of gradient flow smoothing applied before measuring $n$. We conclude that the semiclassical equation (\ref{eq:photon-mass-semiclassical}) is approximately valid even in the crossover region if the monopole number density is renormalized using the gradient flow.

  \section{Conclusions}
  \label{sec:conclusions}
  
  We have studied the crossover transition in the
  three-dimensional $\SU2$ Georgi-Glashow model, with a particular focus on the behavior
  of the 't Hooft-Polyakov magnetic monopoles during the
  transition. A gas of monopoles is present in the Higgs regime.
  Any given lattice configuration therefore has many monopoles, a
  large number of which are due to fluctuations of the gauge field on
  length scales close to the lattice spacing. These monopoles (often
  identifiable as monopole-antimonopole excitations) obscure the
  infrared physics of screening in the monopole gas, which we wished
  to investigate here.

  We employed a gradient flow method to smooth the gauge field
  configurations, removing the short-distance fluctuations, and measured 
  the resulting renormalized number
  density $n$ of monopoles with a given smoothing radius $\xi$. This method was shown to give a continuum limit result for $n$.
  We found that large lattices were required to reach the infinite volume
  limit for the number density. The volume dependence is as expected based on the semiclassical dilute gas description, in which monopoles are screened at distances longer than their mean separation. 

  The screening means that the perturbative ``photon'' is removed from the spectrum and replaced by a massive pseudoscalar. In the present work we have measured this nonzero mass on the lattice 
  near the crossover, directly confirming both sides of Polyakov's 
  picture of the transition. The mass is vanishingly small deep in the Higgs regime.
  Polyakov's semiclassical description of confinement in the presence of monopoles
  yields a relationship between monopole number density and the mass of the
  photon-like excitation. We have confirmed this relation nonperturbatively for our renormalized definition of the number density, up to a constant of proportionality.
  We continued our photon mass operator and the monopole number measurements 
  into the confining ``phase'' and our results suggest that the 
  relationship predicted by Polyakov could persist there as
  well, despite the physical interpretations of the two
  observables being less clear. This could be interesting to study in more detail.

  Although we did not study first-order transitions in this work, we would expect
  the behavior of monopoles to be quite similar. In the crossover scenario 
  studied here, the system interpolates smoothly between the two ``phases'',
  and the photon is unambiguously massive in the intermediate region.
  In contrast, a first-order transition means that the Higgs condensate jumps directly to a large value and suppresses the rate of monopole-antimonopole pair production. Consequently, the photon mass may be too small to detect in such scenarios.

  In summary, our results provide strong evidence that the semiclassical picture of confinement in the three-dimensional $\SU2$ model stands up to nonperturbative
  study. This could help to shed light on the role monopoles play in
  the vicinity of transitions in this model, and other models with similar field content.

  \bigskip

  \noindent
  \textbf{Acknowledgments} We acknowledge useful discussions with
  Arttu Rajantie. 
  This work was supported by Academy of Finland grant nos. 308791, 320123, 324882 and 328958.
  LN acknowledges financial support from the Jenny and Antti Wihuri Foundation.
  We thank Jarno Rantaharju for collaboration in early stages of this work.

  \newpage
  \appendix
  
  \section{Details of the gradient flow}
  \label{sec:grad-flow-appendix}
  
  This appendix collects the required expressions for integrating the gradient flow equations (\ref{eq:grad-flow-1})-(\ref{eq:grad-flow-2}). Consider the lattice action eq.\,(\ref{eq:lattice-action}) and denote $T^a = \frac12 \sigma^a$. The gauge links can be written in terms of Lie algebra fields as $U_i(x) = \exp \left[i T^a \theta_i^a(x) \right]$ and we rescale the Higgs field to natural lattice units, $\hat\phi = \sqrt{a} \phi$.
  
  Our flow equations read, in $d=3$ dimensions,
  \begin{align}
  \frac{\partial U_i(x)}{\partial\tau} &= -i g_3^2 a T^a \frac{\partial S}{\partial \theta_i^a(x)} U_i(x) \equiv Z[U_i(x)] U_i(x) \\
  \frac{\partial \hat\phi^a(x)}{\partial \tau} &= - \frac{\partial S}{\partial \hat\phi^a(x)},
  \end{align}
  where $U_i(x)$ and $\hat\phi(x)$ are implicitly understood to be functions of the flow time $\tau$. The lattice gradients are given by 
  \begin{align}
  \frac{\partial S}{\partial \hat\phi^a(x)} =& \; 6 \hat\phi^a(x) - 2\Tr\Big[ T^a U_i(x) \hat\phi(x+i) \he U_i(x) + T^a \he U_i(x-i) \hat\phi(x-i) U_i(x-i) \Big] \nn
  & + (am_L)^2 \hat\phi^a(x) + (a\lambda_3) \left( 2\Tr\phi(x)^2 \right) \hat\phi^a(x)
  \end{align}
  and
  \begin{align}
  Z[U_i(x)] =& \; -\frac{4}{\beta} \Big[ M - \he M - \frac{1}{2} \Tr \Big( M - \he M \Big) \Big],
  \end{align}
  where 
  \begin{align}
  M &= \; \frac{\beta}{8} U_i(x) \sum_{j \neq i} \Big[ U_j(x+i) \he U_i(x+j) \he U_j(x) + \he U_j(x+i-j) \he U_i(x-j) U_j(x-j) \Big] \nn 
  & - U_i(x) \hat\phi(x+i) \he U_i(x) \hat\phi(x)
  \end{align}
  is the product of $U_i(x)$ and a generalized ``staple'' for the gauge-Higgs system. The normalization of $\tau$ is chosen to match with ref.~\cite{Luscher:2010iy} and from the analysis there one sees that the $d=3$ smoothing radius is $\sqrt{6\tau}a$.

  We integrate the flow using a simple Euler integrator:
  \begin{align}
  U_i(x, \tau + \Delta\tau) &= U_i(x,\tau) e^{ Z[U_i(x,\tau)] \Delta\tau} \\
  \hat\phi^a(x, \tau + \Delta\tau) &= \hat\phi^a(x,\tau) - \frac{\partial S}{\partial \hat\phi^a(x,\tau)} \Delta\tau.
  \end{align}
  This method is sufficient for our purposes; see however Appendix C in \cite{Luscher:2010iy} for a more accurate Runge-Kutta integrator. The time step used in our simulations was $\Delta\tau = 0.025$.


  \bibliography{su2adj}

\begin{thebibliography}{30}%
\makeatletter
\providecommand \@ifxundefined [1]{%
 \@ifx{#1\undefined}
}%
\providecommand \@ifnum [1]{%
 \ifnum #1\expandafter \@firstoftwo
 \else \expandafter \@secondoftwo
 \fi
}%
\providecommand \@ifx [1]{%
 \ifx #1\expandafter \@firstoftwo
 \else \expandafter \@secondoftwo
 \fi
}%
\providecommand \natexlab [1]{#1}%
\providecommand \enquote  [1]{``#1''}%
\providecommand \bibnamefont  [1]{#1}%
\providecommand \bibfnamefont [1]{#1}%
\providecommand \citenamefont [1]{#1}%
\providecommand \href@noop [0]{\@secondoftwo}%
\providecommand \href [0]{\begingroup \@sanitize@url \@href}%
\providecommand \@href[1]{\@@startlink{#1}\@@href}%
\providecommand \@@href[1]{\endgroup#1\@@endlink}%
\providecommand \@sanitize@url [0]{\catcode `\\12\catcode `\$12\catcode
  `\&12\catcode `\#12\catcode `\^12\catcode `\_12\catcode `\%12\relax}%
\providecommand \@@startlink[1]{}%
\providecommand \@@endlink[0]{}%
\providecommand \url  [0]{\begingroup\@sanitize@url \@url }%
\providecommand \@url [1]{\endgroup\@href {#1}{\urlprefix }}%
\providecommand \urlprefix  [0]{URL }%
\providecommand \Eprint [0]{\href }%
\providecommand \doibase [0]{http://dx.doi.org/}%
\providecommand \selectlanguage [0]{\@gobble}%
\providecommand \bibinfo  [0]{\@secondoftwo}%
\providecommand \bibfield  [0]{\@secondoftwo}%
\providecommand \translation [1]{[#1]}%
\providecommand \BibitemOpen [0]{}%
\providecommand \bibitemStop [0]{}%
\providecommand \bibitemNoStop [0]{.\EOS\space}%
\providecommand \EOS [0]{\spacefactor3000\relax}%
\providecommand \BibitemShut  [1]{\csname bibitem#1\endcsname}%
\let\auto@bib@innerbib\@empty
\bibitem [{\citenamefont {Georgi}\ and\ \citenamefont
  {Glashow}(1974)}]{Georgi:1974sy}%
  \BibitemOpen
  \bibfield  {author} {\bibinfo {author} {\bibfnamefont {H.}~\bibnamefont
  {Georgi}}\ and\ \bibinfo {author} {\bibfnamefont {S.~L.}\ \bibnamefont
  {Glashow}},\ }\href {\doibase 10.1103/PhysRevLett.32.438} {\bibfield
  {journal} {\bibinfo  {journal} {Phys. Rev. Lett.}\ }\textbf {\bibinfo
  {volume} {32}},\ \bibinfo {pages} {438} (\bibinfo {year} {1974})}\BibitemShut
  {NoStop}%
\bibitem [{\citenamefont {Appelquist}\ and\ \citenamefont
  {Pisarski}(1981)}]{Appelquist:1981vg}%
  \BibitemOpen
  \bibfield  {author} {\bibinfo {author} {\bibfnamefont {T.}~\bibnamefont
  {Appelquist}}\ and\ \bibinfo {author} {\bibfnamefont {R.~D.}\ \bibnamefont
  {Pisarski}},\ }\href {\doibase 10.1103/PhysRevD.23.2305} {\bibfield
  {journal} {\bibinfo  {journal} {Phys. Rev. D}\ }\textbf {\bibinfo {volume}
  {23}},\ \bibinfo {pages} {2305} (\bibinfo {year} {1981})}\BibitemShut
  {NoStop}%
\bibitem [{\citenamefont {Kajantie}\ \emph {et~al.}(1997)\citenamefont
  {Kajantie}, \citenamefont {Laine}, \citenamefont {Rummukainen},\ and\
  \citenamefont {Shaposhnikov}}]{Kajantie:1997tt}%
  \BibitemOpen
  \bibfield  {author} {\bibinfo {author} {\bibfnamefont {K.}~\bibnamefont
  {Kajantie}}, \bibinfo {author} {\bibfnamefont {M.}~\bibnamefont {Laine}},
  \bibinfo {author} {\bibfnamefont {K.}~\bibnamefont {Rummukainen}}, \ and\
  \bibinfo {author} {\bibfnamefont {M.~E.}\ \bibnamefont {Shaposhnikov}},\
  }\href {\doibase 10.1016/S0550-3213(97)00425-2} {\bibfield  {journal}
  {\bibinfo  {journal} {Nucl. Phys. B}\ }\textbf {\bibinfo {volume} {503}},\
  \bibinfo {pages} {357} (\bibinfo {year} {1997})},\ \Eprint
  {http://arxiv.org/abs/hep-ph/9704416} {arXiv:hep-ph/9704416} \BibitemShut
  {NoStop}%
\bibitem [{\citenamefont {'t~Hooft}(1974)}]{tHooft:1974kcl}%
  \BibitemOpen
  \bibfield  {author} {\bibinfo {author} {\bibfnamefont {G.}~\bibnamefont
  {'t~Hooft}},\ }\href {\doibase 10.1016/0550-3213(74)90486-6} {\bibfield
  {journal} {\bibinfo  {journal} {Nucl. Phys. B}\ }\textbf {\bibinfo {volume}
  {79}},\ \bibinfo {pages} {276} (\bibinfo {year} {1974})}\BibitemShut
  {NoStop}%
\bibitem [{\citenamefont {Polyakov}(1974)}]{Polyakov:1974ek}%
  \BibitemOpen
  \bibfield  {author} {\bibinfo {author} {\bibfnamefont {A.~M.}\ \bibnamefont
  {Polyakov}},\ }\href@noop {} {\bibfield  {journal} {\bibinfo  {journal} {JETP
  Lett.}\ }\textbf {\bibinfo {volume} {20}},\ \bibinfo {pages} {194} (\bibinfo
  {year} {1974})}\BibitemShut {NoStop}%
\bibitem [{\citenamefont {Polyakov}(1977)}]{Polyakov:1976fu}%
  \BibitemOpen
  \bibfield  {author} {\bibinfo {author} {\bibfnamefont {A.~M.}\ \bibnamefont
  {Polyakov}},\ }\href {\doibase 10.1016/0550-3213(77)90086-4} {\bibfield
  {journal} {\bibinfo  {journal} {Nucl. Phys. B}\ }\textbf {\bibinfo {volume}
  {120}},\ \bibinfo {pages} {429} (\bibinfo {year} {1977})}\BibitemShut
  {NoStop}%
\bibitem [{\citenamefont {Nadkarni}(1990)}]{Nadkarni:1989na}%
  \BibitemOpen
  \bibfield  {author} {\bibinfo {author} {\bibfnamefont {S.}~\bibnamefont
  {Nadkarni}},\ }\href {\doibase 10.1016/0550-3213(90)90491-U} {\bibfield
  {journal} {\bibinfo  {journal} {Nucl. Phys. B}\ }\textbf {\bibinfo {volume}
  {334}},\ \bibinfo {pages} {559} (\bibinfo {year} {1990})}\BibitemShut
  {NoStop}%
\bibitem [{\citenamefont {Hart}\ \emph {et~al.}(1997)\citenamefont {Hart},
  \citenamefont {Philipsen}, \citenamefont {Stack},\ and\ \citenamefont
  {Teper}}]{Hart:1996ac}%
  \BibitemOpen
  \bibfield  {author} {\bibinfo {author} {\bibfnamefont {A.}~\bibnamefont
  {Hart}}, \bibinfo {author} {\bibfnamefont {O.}~\bibnamefont {Philipsen}},
  \bibinfo {author} {\bibfnamefont {J.~D.}\ \bibnamefont {Stack}}, \ and\
  \bibinfo {author} {\bibfnamefont {M.}~\bibnamefont {Teper}},\ }\href
  {\doibase 10.1016/S0370-2693(97)00104-4} {\bibfield  {journal} {\bibinfo
  {journal} {Phys. Lett. B}\ }\textbf {\bibinfo {volume} {396}},\ \bibinfo
  {pages} {217} (\bibinfo {year} {1997})},\ \Eprint
  {http://arxiv.org/abs/hep-lat/9612021} {arXiv:hep-lat/9612021} \BibitemShut
  {NoStop}%
\bibitem [{\citenamefont {Davis}\ \emph {et~al.}(2002)\citenamefont {Davis},
  \citenamefont {Hart}, \citenamefont {Kibble},\ and\ \citenamefont
  {Rajantie}}]{Davis:2001mg}%
  \BibitemOpen
  \bibfield  {author} {\bibinfo {author} {\bibfnamefont {A.~C.}\ \bibnamefont
  {Davis}}, \bibinfo {author} {\bibfnamefont {A.}~\bibnamefont {Hart}},
  \bibinfo {author} {\bibfnamefont {T.~W.~B.}\ \bibnamefont {Kibble}}, \ and\
  \bibinfo {author} {\bibfnamefont {A.}~\bibnamefont {Rajantie}},\ }\href
  {\doibase 10.1103/PhysRevD.65.125008} {\bibfield  {journal} {\bibinfo
  {journal} {Phys. Rev. D}\ }\textbf {\bibinfo {volume} {65}},\ \bibinfo
  {pages} {125008} (\bibinfo {year} {2002})},\ \Eprint
  {http://arxiv.org/abs/hep-lat/0110154} {arXiv:hep-lat/0110154} \BibitemShut
  {NoStop}%
\bibitem [{\citenamefont {Lee}\ and\ \citenamefont
  {Shigemitsu}(1986)}]{Lee:1985yi}%
  \BibitemOpen
  \bibfield  {author} {\bibinfo {author} {\bibfnamefont {I.-H.}\ \bibnamefont
  {Lee}}\ and\ \bibinfo {author} {\bibfnamefont {J.}~\bibnamefont
  {Shigemitsu}},\ }\href {\doibase 10.1016/0550-3213(86)90117-3} {\bibfield
  {journal} {\bibinfo  {journal} {Nucl. Phys. B}\ }\textbf {\bibinfo {volume}
  {263}},\ \bibinfo {pages} {280} (\bibinfo {year} {1986})}\BibitemShut
  {NoStop}%
\bibitem [{\citenamefont {Afferrante}\ \emph {et~al.}(2020)\citenamefont
  {Afferrante}, \citenamefont {Maas},\ and\ \citenamefont
  {T\"orek}}]{Afferrante:2020hqe}%
  \BibitemOpen
  \bibfield  {author} {\bibinfo {author} {\bibfnamefont {V.}~\bibnamefont
  {Afferrante}}, \bibinfo {author} {\bibfnamefont {A.}~\bibnamefont {Maas}}, \
  and\ \bibinfo {author} {\bibfnamefont {P.}~\bibnamefont {T\"orek}},\ }\href
  {\doibase 10.1103/PhysRevD.101.114506} {\bibfield  {journal} {\bibinfo
  {journal} {Phys. Rev. D}\ }\textbf {\bibinfo {volume} {101}},\ \bibinfo
  {pages} {114506} (\bibinfo {year} {2020})},\ \Eprint
  {http://arxiv.org/abs/2002.08221} {arXiv:2002.08221 [hep-lat]} \BibitemShut
  {NoStop}%
\bibitem [{\citenamefont {L\"uscher}(2010)}]{Luscher:2010iy}%
  \BibitemOpen
  \bibfield  {author} {\bibinfo {author} {\bibfnamefont {M.}~\bibnamefont
  {L\"uscher}},\ }\href {\doibase 10.1007/JHEP08(2010)071} {\bibfield
  {journal} {\bibinfo  {journal} {JHEP}\ }\textbf {\bibinfo {volume} {08}},\
  \bibinfo {pages} {071} (\bibinfo {year} {2010})},\ \bibinfo {note} {[Erratum:
  JHEP 03, 092 (2014)]},\ \Eprint {http://arxiv.org/abs/1006.4518}
  {arXiv:1006.4518 [hep-lat]} \BibitemShut {NoStop}%
\bibitem [{\citenamefont {Niemi}\ \emph
  {et~al.}(2021{\natexlab{a}})\citenamefont {Niemi}, \citenamefont
  {Rummukainen}, \citenamefont {Sepp\"a},\ and\ \citenamefont
  {Weir}}]{Niemi:2021ghk}%
  \BibitemOpen
  \bibfield  {author} {\bibinfo {author} {\bibfnamefont {L.}~\bibnamefont
  {Niemi}}, \bibinfo {author} {\bibfnamefont {K.}~\bibnamefont {Rummukainen}},
  \bibinfo {author} {\bibfnamefont {R.}~\bibnamefont {Sepp\"a}}, \ and\
  \bibinfo {author} {\bibfnamefont {D.}~\bibnamefont {Weir}},\ }in\ \href@noop
  {} {\emph {\bibinfo {booktitle} {{38th International Symposium on Lattice
  Field Theory}}}}\ (\bibinfo {year} {2021})\ \Eprint
  {http://arxiv.org/abs/2111.09097} {arXiv:2111.09097 [hep-lat]} \BibitemShut
  {NoStop}%
\bibitem [{\citenamefont {Patel}\ and\ \citenamefont
  {Ramsey-Musolf}(2013)}]{Patel:2012pi}%
  \BibitemOpen
  \bibfield  {author} {\bibinfo {author} {\bibfnamefont {H.~H.}\ \bibnamefont
  {Patel}}\ and\ \bibinfo {author} {\bibfnamefont {M.~J.}\ \bibnamefont
  {Ramsey-Musolf}},\ }\href {\doibase 10.1103/PhysRevD.88.035013} {\bibfield
  {journal} {\bibinfo  {journal} {Phys. Rev. D}\ }\textbf {\bibinfo {volume}
  {88}},\ \bibinfo {pages} {035013} (\bibinfo {year} {2013})},\ \Eprint
  {http://arxiv.org/abs/1212.5652} {arXiv:1212.5652 [hep-ph]} \BibitemShut
  {NoStop}%
\bibitem [{\citenamefont {Niemi}\ \emph
  {et~al.}(2021{\natexlab{b}})\citenamefont {Niemi}, \citenamefont
  {Ramsey-Musolf}, \citenamefont {Tenkanen},\ and\ \citenamefont
  {Weir}}]{Niemi:2020hto}%
  \BibitemOpen
  \bibfield  {author} {\bibinfo {author} {\bibfnamefont {L.}~\bibnamefont
  {Niemi}}, \bibinfo {author} {\bibfnamefont {M.~J.}\ \bibnamefont
  {Ramsey-Musolf}}, \bibinfo {author} {\bibfnamefont {T.~V.~I.}\ \bibnamefont
  {Tenkanen}}, \ and\ \bibinfo {author} {\bibfnamefont {D.~J.}\ \bibnamefont
  {Weir}},\ }\href {\doibase 10.1103/PhysRevLett.126.171802} {\bibfield
  {journal} {\bibinfo  {journal} {Phys. Rev. Lett.}\ }\textbf {\bibinfo
  {volume} {126}},\ \bibinfo {pages} {171802} (\bibinfo {year}
  {2021}{\natexlab{b}})},\ \Eprint {http://arxiv.org/abs/2005.11332}
  {arXiv:2005.11332 [hep-ph]} \BibitemShut {NoStop}%
\bibitem [{\citenamefont {McFadden}\ and\ \citenamefont
  {Skenderis}(2010)}]{McFadden:2009fg}%
  \BibitemOpen
  \bibfield  {author} {\bibinfo {author} {\bibfnamefont {P.}~\bibnamefont
  {McFadden}}\ and\ \bibinfo {author} {\bibfnamefont {K.}~\bibnamefont
  {Skenderis}},\ }\href {\doibase 10.1103/PhysRevD.81.021301} {\bibfield
  {journal} {\bibinfo  {journal} {Phys. Rev. D}\ }\textbf {\bibinfo {volume}
  {81}},\ \bibinfo {pages} {021301} (\bibinfo {year} {2010})},\ \Eprint
  {http://arxiv.org/abs/0907.5542} {arXiv:0907.5542 [hep-th]} \BibitemShut
  {NoStop}%
\bibitem [{\citenamefont {Kajantie}\ \emph {et~al.}(1999)\citenamefont
  {Kajantie}, \citenamefont {Laine}, \citenamefont {Neuhaus}, \citenamefont
  {Peisa}, \citenamefont {Rajantie},\ and\ \citenamefont
  {Rummukainen}}]{Kajantie:1998zn}%
  \BibitemOpen
  \bibfield  {author} {\bibinfo {author} {\bibfnamefont {K.}~\bibnamefont
  {Kajantie}}, \bibinfo {author} {\bibfnamefont {M.}~\bibnamefont {Laine}},
  \bibinfo {author} {\bibfnamefont {T.}~\bibnamefont {Neuhaus}}, \bibinfo
  {author} {\bibfnamefont {J.}~\bibnamefont {Peisa}}, \bibinfo {author}
  {\bibfnamefont {A.}~\bibnamefont {Rajantie}}, \ and\ \bibinfo {author}
  {\bibfnamefont {K.}~\bibnamefont {Rummukainen}},\ }\href {\doibase
  10.1016/S0550-3213(99)00033-4} {\bibfield  {journal} {\bibinfo  {journal}
  {Nucl. Phys. B}\ }\textbf {\bibinfo {volume} {546}},\ \bibinfo {pages} {351}
  (\bibinfo {year} {1999})},\ \Eprint {http://arxiv.org/abs/hep-ph/9809334}
  {arXiv:hep-ph/9809334} \BibitemShut {NoStop}%
\bibitem [{\citenamefont {Forgacs}\ \emph {et~al.}(2005)\citenamefont
  {Forgacs}, \citenamefont {Obadia},\ and\ \citenamefont
  {Reuillon}}]{Forgacs:2005vx}%
  \BibitemOpen
  \bibfield  {author} {\bibinfo {author} {\bibfnamefont {P.}~\bibnamefont
  {Forgacs}}, \bibinfo {author} {\bibfnamefont {N.}~\bibnamefont {Obadia}}, \
  and\ \bibinfo {author} {\bibfnamefont {S.}~\bibnamefont {Reuillon}},\ }\href
  {\doibase 10.1103/PhysRevD.71.035002} {\bibfield  {journal} {\bibinfo
  {journal} {Phys. Rev. D}\ }\textbf {\bibinfo {volume} {71}},\ \bibinfo
  {pages} {035002} (\bibinfo {year} {2005})},\ \bibinfo {note} {[Erratum:
  Phys.Rev.D 71, 119902 (2005)]},\ \Eprint
  {http://arxiv.org/abs/hep-th/0412057} {arXiv:hep-th/0412057} \BibitemShut
  {NoStop}%
\bibitem [{\citenamefont {Laine}(1995)}]{Laine:1995np}%
  \BibitemOpen
  \bibfield  {author} {\bibinfo {author} {\bibfnamefont {M.}~\bibnamefont
  {Laine}},\ }\href {\doibase 10.1016/0550-3213(95)00356-W} {\bibfield
  {journal} {\bibinfo  {journal} {Nucl. Phys. B}\ }\textbf {\bibinfo {volume}
  {451}},\ \bibinfo {pages} {484} (\bibinfo {year} {1995})},\ \Eprint
  {http://arxiv.org/abs/hep-lat/9504001} {arXiv:hep-lat/9504001} \BibitemShut
  {NoStop}%
\bibitem [{\citenamefont {Laine}\ and\ \citenamefont
  {Rajantie}(1998)}]{Laine:1997dy}%
  \BibitemOpen
  \bibfield  {author} {\bibinfo {author} {\bibfnamefont {M.}~\bibnamefont
  {Laine}}\ and\ \bibinfo {author} {\bibfnamefont {A.}~\bibnamefont
  {Rajantie}},\ }\href {\doibase 10.1016/S0550-3213(97)00709-8} {\bibfield
  {journal} {\bibinfo  {journal} {Nucl. Phys. B}\ }\textbf {\bibinfo {volume}
  {513}},\ \bibinfo {pages} {471} (\bibinfo {year} {1998})},\ \Eprint
  {http://arxiv.org/abs/hep-lat/9705003} {arXiv:hep-lat/9705003} \BibitemShut
  {NoStop}%
\bibitem [{\citenamefont {Moore}(1998)}]{Moore:1997np}%
  \BibitemOpen
  \bibfield  {author} {\bibinfo {author} {\bibfnamefont {G.~D.}\ \bibnamefont
  {Moore}},\ }\href {\doibase 10.1016/S0550-3213(98)00158-8} {\bibfield
  {journal} {\bibinfo  {journal} {Nucl. Phys. B}\ }\textbf {\bibinfo {volume}
  {523}},\ \bibinfo {pages} {569} (\bibinfo {year} {1998})},\ \Eprint
  {http://arxiv.org/abs/hep-lat/9709053} {arXiv:hep-lat/9709053} \BibitemShut
  {NoStop}%
\bibitem [{\citenamefont {Moore}\ and\ \citenamefont
  {Schlusser}(2019)}]{Moore:2019lua}%
  \BibitemOpen
  \bibfield  {author} {\bibinfo {author} {\bibfnamefont {G.~D.}\ \bibnamefont
  {Moore}}\ and\ \bibinfo {author} {\bibfnamefont {N.}~\bibnamefont
  {Schlusser}},\ }\href {\doibase 10.1103/PhysRevD.100.034510} {\bibfield
  {journal} {\bibinfo  {journal} {Phys. Rev. D}\ }\textbf {\bibinfo {volume}
  {100}},\ \bibinfo {pages} {034510} (\bibinfo {year} {2019})},\ \Eprint
  {http://arxiv.org/abs/1905.09708} {arXiv:1905.09708 [hep-lat]} \BibitemShut
  {NoStop}%
\bibitem [{\citenamefont {Davis}\ \emph {et~al.}(2000)\citenamefont {Davis},
  \citenamefont {Kibble}, \citenamefont {Rajantie},\ and\ \citenamefont
  {Shanahan}}]{Davis:2000kv}%
  \BibitemOpen
  \bibfield  {author} {\bibinfo {author} {\bibfnamefont {A.~C.}\ \bibnamefont
  {Davis}}, \bibinfo {author} {\bibfnamefont {T.~W.~B.}\ \bibnamefont
  {Kibble}}, \bibinfo {author} {\bibfnamefont {A.}~\bibnamefont {Rajantie}}, \
  and\ \bibinfo {author} {\bibfnamefont {H.}~\bibnamefont {Shanahan}},\ }\href
  {\doibase 10.1088/1126-6708/2000/11/010} {\bibfield  {journal} {\bibinfo
  {journal} {JHEP}\ }\textbf {\bibinfo {volume} {11}},\ \bibinfo {pages} {010}
  (\bibinfo {year} {2000})},\ \Eprint {http://arxiv.org/abs/hep-lat/0009037}
  {arXiv:hep-lat/0009037} \BibitemShut {NoStop}%
\bibitem [{\citenamefont {Kronfeld}\ and\ \citenamefont
  {Wiese}(1991)}]{Kronfeld:1990qu}%
  \BibitemOpen
  \bibfield  {author} {\bibinfo {author} {\bibfnamefont {A.~S.}\ \bibnamefont
  {Kronfeld}}\ and\ \bibinfo {author} {\bibfnamefont {U.~J.}\ \bibnamefont
  {Wiese}},\ }\href {\doibase 10.1016/0550-3213(91)90479-H} {\bibfield
  {journal} {\bibinfo  {journal} {Nucl. Phys. B}\ }\textbf {\bibinfo {volume}
  {357}},\ \bibinfo {pages} {521} (\bibinfo {year} {1991})}\BibitemShut
  {NoStop}%
\bibitem [{\citenamefont {Edwards}\ \emph {et~al.}(2009)\citenamefont
  {Edwards}, \citenamefont {Mehta}, \citenamefont {Rajantie},\ and\
  \citenamefont {von Smekal}}]{Edwards:2009bw}%
  \BibitemOpen
  \bibfield  {author} {\bibinfo {author} {\bibfnamefont {S.}~\bibnamefont
  {Edwards}}, \bibinfo {author} {\bibfnamefont {D.~B.}\ \bibnamefont {Mehta}},
  \bibinfo {author} {\bibfnamefont {A.}~\bibnamefont {Rajantie}}, \ and\
  \bibinfo {author} {\bibfnamefont {L.}~\bibnamefont {von Smekal}},\ }\href
  {\doibase 10.1103/PhysRevD.80.065030} {\bibfield  {journal} {\bibinfo
  {journal} {Phys. Rev. D}\ }\textbf {\bibinfo {volume} {80}},\ \bibinfo
  {pages} {065030} (\bibinfo {year} {2009})},\ \Eprint
  {http://arxiv.org/abs/0906.5531} {arXiv:0906.5531 [hep-lat]} \BibitemShut
  {NoStop}%
\bibitem [{\citenamefont {Rajantie}(2006)}]{Rajantie:2005hi}%
  \BibitemOpen
  \bibfield  {author} {\bibinfo {author} {\bibfnamefont {A.}~\bibnamefont
  {Rajantie}},\ }\href {\doibase 10.1088/1126-6708/2006/01/088} {\bibfield
  {journal} {\bibinfo  {journal} {JHEP}\ }\textbf {\bibinfo {volume} {01}},\
  \bibinfo {pages} {088} (\bibinfo {year} {2006})},\ \Eprint
  {http://arxiv.org/abs/hep-lat/0512006} {arXiv:hep-lat/0512006} \BibitemShut
  {NoStop}%
\bibitem [{\citenamefont {Moore}(1999)}]{Moore:1998swa}%
  \BibitemOpen
  \bibfield  {author} {\bibinfo {author} {\bibfnamefont {G.~D.}\ \bibnamefont
  {Moore}},\ }\href {\doibase 10.1103/PhysRevD.59.014503} {\bibfield  {journal}
  {\bibinfo  {journal} {Phys. Rev. D}\ }\textbf {\bibinfo {volume} {59}},\
  \bibinfo {pages} {014503} (\bibinfo {year} {1999})},\ \Eprint
  {http://arxiv.org/abs/hep-ph/9805264} {arXiv:hep-ph/9805264} \BibitemShut
  {NoStop}%
\bibitem [{\citenamefont {Kennedy}\ and\ \citenamefont
  {Pendleton}(1985)}]{Kennedy:1985nu}%
  \BibitemOpen
  \bibfield  {author} {\bibinfo {author} {\bibfnamefont {A.~D.}\ \bibnamefont
  {Kennedy}}\ and\ \bibinfo {author} {\bibfnamefont {B.~J.}\ \bibnamefont
  {Pendleton}},\ }\href {\doibase 10.1016/0370-2693(85)91632-6} {\bibfield
  {journal} {\bibinfo  {journal} {Phys. Lett. B}\ }\textbf {\bibinfo {volume}
  {156}},\ \bibinfo {pages} {393} (\bibinfo {year} {1985})}\BibitemShut
  {NoStop}%
\bibitem [{\citenamefont {Kajantie}\ \emph {et~al.}(1996)\citenamefont
  {Kajantie}, \citenamefont {Laine}, \citenamefont {Rummukainen},\ and\
  \citenamefont {Shaposhnikov}}]{Kajantie:1995kf}%
  \BibitemOpen
  \bibfield  {author} {\bibinfo {author} {\bibfnamefont {K.}~\bibnamefont
  {Kajantie}}, \bibinfo {author} {\bibfnamefont {M.}~\bibnamefont {Laine}},
  \bibinfo {author} {\bibfnamefont {K.}~\bibnamefont {Rummukainen}}, \ and\
  \bibinfo {author} {\bibfnamefont {M.~E.}\ \bibnamefont {Shaposhnikov}},\
  }\href {\doibase 10.1016/0550-3213(96)00052-1} {\bibfield  {journal}
  {\bibinfo  {journal} {Nucl. Phys. B}\ }\textbf {\bibinfo {volume} {466}},\
  \bibinfo {pages} {189} (\bibinfo {year} {1996})},\ \Eprint
  {http://arxiv.org/abs/hep-lat/9510020} {arXiv:hep-lat/9510020} \BibitemShut
  {NoStop}%
\bibitem [{\citenamefont {Bonnet}\ \emph {et~al.}(2002)\citenamefont {Bonnet},
  \citenamefont {Leinweber}, \citenamefont {Williams},\ and\ \citenamefont
  {Zanotti}}]{Bonnet:2001rc}%
  \BibitemOpen
  \bibfield  {author} {\bibinfo {author} {\bibfnamefont {F.~D.~R.}\
  \bibnamefont {Bonnet}}, \bibinfo {author} {\bibfnamefont {D.~B.}\
  \bibnamefont {Leinweber}}, \bibinfo {author} {\bibfnamefont {A.~G.}\
  \bibnamefont {Williams}}, \ and\ \bibinfo {author} {\bibfnamefont {J.~M.}\
  \bibnamefont {Zanotti}},\ }\href {\doibase 10.1103/PhysRevD.65.114510}
  {\bibfield  {journal} {\bibinfo  {journal} {Phys. Rev. D}\ }\textbf {\bibinfo
  {volume} {65}},\ \bibinfo {pages} {114510} (\bibinfo {year} {2002})},\
  \Eprint {http://arxiv.org/abs/hep-lat/0106023} {arXiv:hep-lat/0106023}
  \BibitemShut {NoStop}%
\end{thebibliography}%

\end{document}